\documentclass[aps,twocolumn,prb,showpacs,10pt,floatfix,groupedaddress]{revtex4-1}
\usepackage{amssymb}
\usepackage{amsmath}
\usepackage{graphicx}
\usepackage{braket}
\usepackage{dcolumn}
\usepackage{bm}
\usepackage{textcomp}
\usepackage{color}
\usepackage{braket}
\usepackage{amsthm}
\usepackage{url}
\usepackage{mathtools}

\begin{document}

\title{Topological phase-diagram of time-periodically rippled zigzag graphene nanoribbons }

\author{Pedro Roman-Taboada}  
\email{peter89@estudiantes.fisica.unam.mx}
\author{Gerardo G. Naumis}

\affiliation{Departamento de Sistemas Complejos, Instituto de
F\'{i}sica, Universidad Nacional Aut\'{o}noma de M\'{e}xico (UNAM),
Apartado Postal 20-364, 01000 M\'{e}xico, Distrito Federal,
M\'{e}xico}

%\date{\today}

\begin{abstract}
The topological properties of electronic edge states in time-periodically driven spatially-periodic corrugated zigzag graphene nanoribbons are studied. An effective 
one-dimensional Hamiltonian is used to describe the electronic properties of graphene and the time-dependence is studied within the Floquet formalism. 
Then the quasienergy spectrum of the evolution operator is obtained using analytical and numeric calculations, both in excellent agreement. Depending on the external parameters of the time-driving,
two different kinds (type I and type II) of touching band points are found, which have a Dirac-like nature at both zero and $\pm\pi$ quasienergy. These touching band points are able to host topologically protected edge states for a finite size system. The topological nature of such edge states was confirmed by an explicit evaluation of the Berry phase in the neighborhood of type I touching band points and by obtaining the winding number of the effective Hamiltonian for type II touching band points. Additionally, the topological phase diagram in terms of the driving parameters of the system was built.
\end{abstract}

%\pacs{73.22.Pr,71.23.Ft,03.65.Vf}

\maketitle

%%%
\section{Introduction}
\label{intro}
%%%

Graphene, a truly two-dimensional material, has proven to have very interesting and fascinating properties\cite{katsnelson07,Review10}. Among them, one can mention its extraordinary mechanical features, which can be used to tailor the electronic properties of graphene and have given rise to many novel effects in the static case\cite{Carrillo14,Nosotros14,Olivaany14,Maurice,wang2015generalized,bahamon2015conductance,roman15,Salary16,EffOliva16,Amorim16,Carrillo16,hernandez2016light,sattari2016spin,sattari2016effects,stegmann2016current,lopez2016magnetic,mikkel2016quantum,ChenSi16,settnes2016pseudomagnetic,Review17,diniz2017graphene-based,Akinwande17,Ghahari17,milovanovic2017graphene,prabhakar2017strain,cariglia2017curvature,bordag2017casimir,zhang2017frequency,nguyen2017optical,Review17}. { As a matter of fact, within the tight binding approach and in the absence of interactions between electrons, the effects of a deformation field applied to graphene can be described via a pseudo magnetic field\cite{Suzuura02,Morpurgo06,Manes07,Castro09,Review17,oliva2017low-energy,oliva2017magento-optical}.} {On the other hand, graphene possesses interesting topological properties for both the time-independent\cite{Vozmediano10,ZakPhase11,Nosotros214,chou2014chalker,Zyuzin2015,gaussi2015zero-field,Mishra2015,Majgraphene15,iorio2015revisting,dal2015line,qu2016electronic,frank17,cao2017topological,wu2017fermionic,wang2017strain} and the time-dependent cases\cite{Delplace13,Thomas14,Gumbs14,perez2014floquet,Usaj14,gavensky16,manghi17,lago17,roman2017topological}. For instance, in the static case, it has been proven that Dirac cones have a non-vanishing Berry phase, which means that they are robust against low perturbation and disorder\cite{Novoselov05}. In addition, since Dirac cones always come in pairs, each cone has an opposite Berry phase as is companion. Hence, as a consequence of the bulk-edge correspondence, an edge state (flat band for the case of pristine zigzag graphene nanoribbons of finite size) emerges joining two inequivalent Dirac cones (this is, two Dirac cones with opposite Berry phase). 

On the other hand, by applying a time-dependent deformation field to graphene, new and novel phenomena appear when compared to the static case \cite{roman2017topological}.} {For instance, when a time-dependent in-plane AC electric field is applied to graphene, it is possible to undergo a topological phase transition from a topological semi-metallic phase to a trivial insulator one\cite{Delplace13merging}. Similarly, gaps on the energy spectrum of graphene can be opened by irradiating graphene with a laser by changing its intensity \cite{Lopez2008,Lopez2010}. This gapped phase is also able to host robust topological chiral Floquet edge states, which are highly tunable\cite{perez2014floquet}. These features are similar to the ones observed in topological insulators, which also exhibit robust edge states. However, there is another kind of topological phases akin to gapless systems\cite{Volovik2011,Volovik2013}. Take the kicked Harper model\cite{KHM16} and the kicked SSH model\cite{wang2017linenodes}, for instance. In the kicked Harper model via periodic driving, one can create many touching band points ({\it i.e.} points at where the band edges touch each other following a linear dispersion) that can give rise to highly localized edge states. This occurs because touching band points always come in pairs and each of them have opposite chirality as its companion\cite{KHM16}. These edge states can be flat bands or dispersive edge states. Interestingly enough, one can have the same effect on graphene nanoribbons by applying a time-dependent strain field\cite{roman2017topological}.} The aim of this paper is to show some of these topological properties of gapless systems by studying a periodically driven uniaxial rippled zigzag graphene nanoribbon (ZGN). To do that, we use a tight-binding Hamiltonian to describe the electronic properties of the periodically driven rippled ZGN within the Floquet formalism. The quasienergy spectrum is then obtained by using an effective Hamiltonian approach.

It is important to remark that the considered deformation field is a corrugation of the graphene ribbon. Here we will restrict ourselves to the case of uniaxial ripple,  i.e., only the height of carbon atoms with respect to a plane is affected only along one direction (in what follows, we will consider a deformation field applied along the armchair direction). Therefore, it is necessary to take into account the relative change of the orientation between $\pi$ orbitals\cite{roman15}.  Within such approximation, as will be seen later on, the time-dependent deformation field allows us to create touching band points (touching band points are points at where a band inversion is observed) at zero or $\pm\pi$ quasienergies. Around such points the quasienergy spectrum exhibits a linear dispersion, as in the case of Dirac cones. The touching band points originated from the time-dependent deformation field can be of two different kinds: type I and type II, each of them giving rise to topologically protected edge states. For the former type, we have found topologically protected flat bands at zero and $\pm\pi$ quasienergy. Such flat bands join two inequivalent touching band points with opposite Berry phase. For the latter, dispersive edge states were found and it was found that they are, at least, topologically weak by obtaining the winding of the effective Hamiltonian.

To finish, it is worthwhile to say that the experimental realization of the deformation pattern here considered can be difficult, since it requires very specific hopping parameters values and because very fast time scales are involved. In fact, a similar experiment  was proposed by us in a previous work \cite{roman2017topological}. {  However, this experiment was tailored for in-plane strain\cite{roman2017topological}, and since graphene is almost incompressible, the compressive strain will induce ripples on the nanoribbon. As a result, it is clear that ripple effects are important to be studied.
Also, it is possible to have a one-dimensional periodic ripple on graphene. This is done by using thermal engineering, and the anisotropic strain pattern can be induced by growing graphene upon a substrate\cite{bai2014creating}. The time-dependent deformation field can be obtained by  applying a time-periodic pressure variation to the whole system\cite{agarwala2016effects,roman2017topological} (the graphene nanoribbon and the substrate). To observe the results presented below, the pressure needs to be in the frequency range of femto seconds, which can be very challenging in a real experiment. As an alternative, we propose the use of artificial or optical lattices to the experimental realization of our model, where the hopping parameters of the graphene nanoribbon lattice can be tailored at will\cite{Uehlinger13,mei2013graphene,Feilhauer15,weinberg16,flaschner2016experimental,dautova2017gapless}. }

The paper is organized as follows, in section \ref{model} we introduce the model. This is, we briefly discuss how to describe the electronic properties of a rippled zigzag graphene nanoribbon. Then, the time dependence is introduced to the model and the time-evolution operator of the system is defined. In section \ref{TBP}, we analytically obtain the quasienergy spectrum of the system via an effective Hamiltonian approach. Also, the location of both types of touching band points is found and the topological phase diagram of the system is built.
The edge states of the system and their topological properties are analyzed in section \ref{edgestates}. Some conclusion are given in section \ref{conclusion}. Finally, in the appendices some calculations regarding the main text are presented. 
\section{Periodically driven rippled graphene}
\label{model}
%

%%%%%%
\begin{figure}
\includegraphics[scale=0.39]{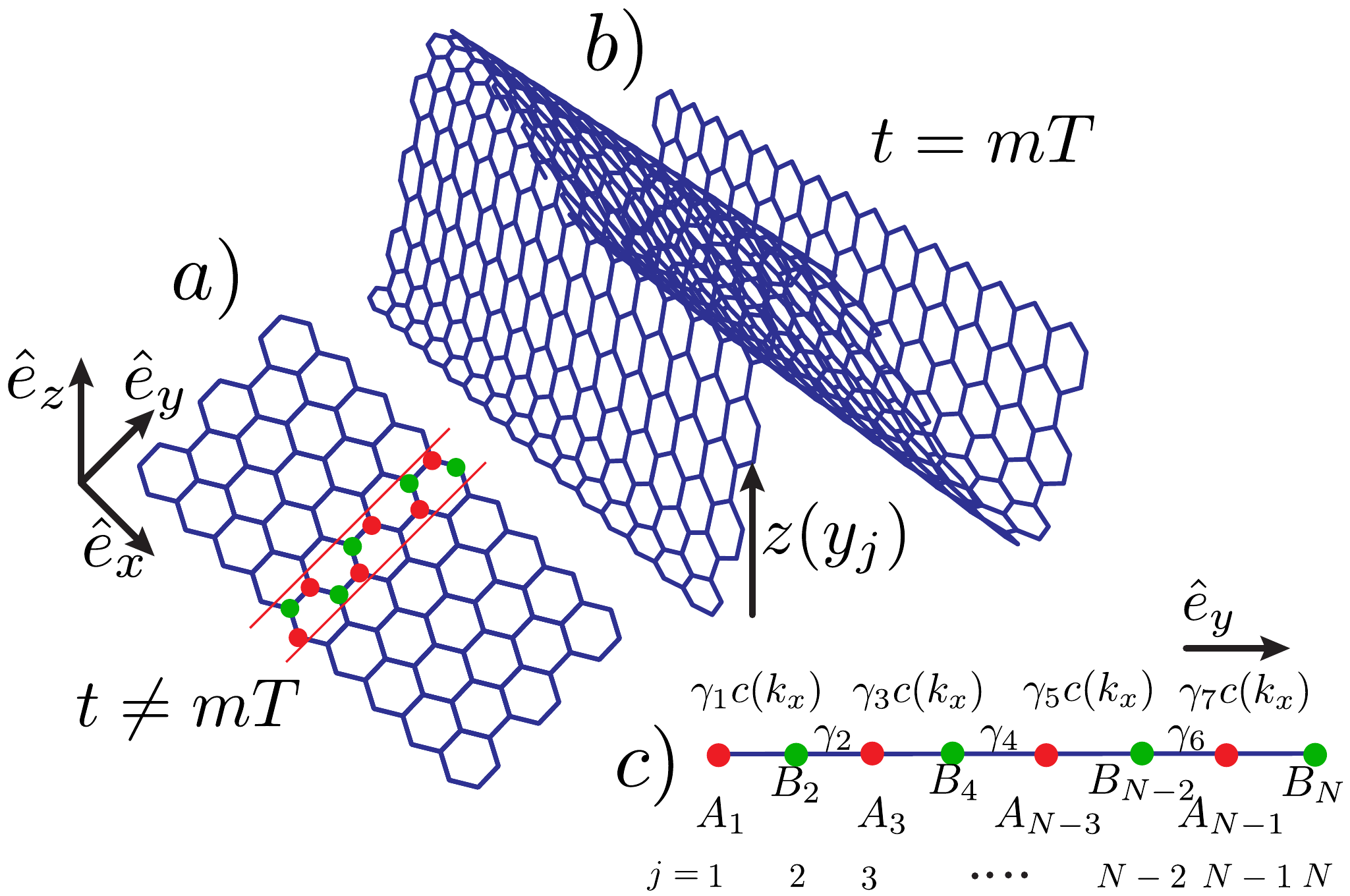}
\caption{(Color online). Schematic representation of the driving layout. The deformation field is turned off for $t\neq mT$, where $t$ is the time, $m$ an integer number, and $T$ is the driving period. This situation is shown in panel a), there in, it can be seen a pristine zigzag graphene nanoribbon (ZGN), which is finite along the $y$-direction but is infinite along the $x$-direction. The unit cell of which is indicated by solid red lines. Atoms belonging to the sub lattice A (B) are indicated by red (green) circles. On the other hand, for $t=mT$ the deformation field is turned on, see panel b). Note that the distance between carbon atoms remains the same as in pristine graphene but the height of each atom is modified along the $y$-direction, such height is given by a spatially periodic function, $z(y)$. Finally, both the pristine and deformed ZGNs can be mapped onto a quasi one-dimensional chain. The mapping of the rippled ZGN is presented in panel c), therein, the same color code as in b) is used. The hopping parameters between carbon atoms are denoted by $\gamma_j$, where $j$ is the label that enumerates the carbon atoms along the $y$-direction within the unit cell. $c(k_x)$ is a function of the quasi-momentum along the $x$-direction, defined in the main text.}
\label{layout}
\end{figure}
%%%%%%

We start by considering a zigzag graphene nanoribbon (ZGN) as the one portrayed in Fig. \ref{layout} a), then we apply an out-of-plane uniaxial deformation field (a ripple field) along the $y$-direction given by,
\begin{equation}
z_{j}=z(y_j)=\lambda\cos{(2\pi\sigma y_j+\phi)},
\label{zdef}
\end{equation}
here, $y_j$ are the positions of the carbon atoms along the $y$-direction, $\lambda$ is the amplitude, $\sigma$ the wavelength, and $\phi$ the phase of the deformation field. Since such a deformation field modifies the height of the carbon atoms, their positions are also modified and can be written as $\mathbf{r^{\prime}}=(\mathbf{r},z(y_j))$, where $\mathbf{r}$ are the carbon atom positions in unrippled graphene. Within the low energy limit, the electronic properties of a zigzag graphene nanoribbon under a deformation field along the armchair direction, as the one given by Eq. (\ref{zdef}), are well described by the following one-dimensional (1D) tight-binding effective Hamiltonian\cite{roman15},
\begin{equation}
H(k_x)=\sum^{N-1}_{j=1} \left[\gamma_{2j}\,a_{2j+1}^{\dag} b_{2j}+c(k_x)\, \gamma_{2j-1} a_{2j-1}^{\dag} b_{2j}\right]+\mathrm{h.c.},
\label{1DHam}
\end{equation}
where $c(k_x)=2\cos{(\sqrt{3}k_x/2)}$, the operator $a_j$ ($b_j$) annihilates an electron at the $j$-th site in the sub lattice A (B), and $N$ is the number of atoms per unit cell (see Fig. \ref{layout}, at where the unit cell is indicated by dotted red lines). $\gamma_{j}$ are the hopping parameters given by\cite{roman15},
\begin{equation}
\gamma_j=\gamma_0\left[1+\alpha \left (1-\hat{\mathbf{N}}_{j}\cdot \hat{\mathbf{N}}_{j+1}\right)\right]\exp{\left(-\beta\delta l_{j+1,j}\right)},
\label{hop}
\end{equation}
where $\gamma_0=2.7$ eV is the hopping parameter for pristine graphene, $\hat{\mathbf{N}}_j$ is the unit vector normal to the pristine graphene sheet at site $j$, which has the following form,
\begin{equation}
\hat{\mathbf{N}}_j=\frac{\hat{e}_z-\nabla z_j}{\sqrt{1+\left(\nabla z_j\right)^2}},
\end{equation}
with $\nabla=(\partial_x,\partial_y)$ being the two-dimensional gradient operator. $\hat{e}_z$ is a unit vector that is perpendicular to the unrippled graphene sheet, $\alpha\approx 0.4$ is a constant that takes into account the change of the relative orientation between $\pi$-orbitals originated from the deformation field, and $\beta\approx3.37$ is the decay rate (Gr\"uneisen parameter). Finally, the quantity $\delta l_{j+1,j}$ is given by,
\begin{equation}
\delta l_{j+1,j}=-1+\sqrt{1+\left[z(y_{j+1})-z(y_j)\right]^2}.
\end{equation}
It is important to say that all distances, here on, will be measured in units of the interatomic distance between carbon atoms ($a_c$) in pristine graphene. In a similar way, we will set $\gamma_0$ as the unit of energy. Having said that, it is noteworthy that the energy spectrum of the Hamiltonian Eq. (\ref{1DHam}) have been discussed in a previous work for the small amplitude limit and for different ripple's wavelength, see reference \cite{roman15}. {  Also, it is important to say that the deformation field here considered induces a pseudo magnetic field, since such deformation field modifies the relative orientation between $\pi$ orbitals. In fact, if we assume that $\mathbf{N}$ is a smooth function of the position, the magnetic flux through a ripple of lateral dimension $l$ and height $z$ is given by\cite{Castro09},
\begin{equation}
\Phi\approx \frac{10\text{\r{A}}^{-1}a_c^2z^2}{l^3}
\end{equation}
If we introduce all the numerical values, we obtain $\Phi\approx10^{-3}\Phi_0$, where $\Phi_0=2\pi\hbar/c$ and $c$ is the speed of light.
}

Once that the Hamiltonian that describes a uniaxial rippled ZGN has been presented, we proceed to introduce the time-dependence to our model. { We will consider a pulse time-driving layout,
\begin{equation}
H(k_x,t)= \left\{\begin{array}{lll}
             H_0(k_x) & if & t<\text{mod}(t,T)<t_1\\
             H_1(k_x)& if & t_1<\text{mod}(t,T)<T
            \end{array}\right.
            \label{shortpulse}
\end{equation}
where $T$ is the driving period and $t_1$ is a number such that $0<t_1<T$. The previous Hamiltonian describes a driving layout in which for times within the interval $(t_1,T)$, the deformation field is turned on, whereas it is turned off for times within the interval $(0,t_1)$. For the sake of simplicity, in what follows we will consider the case of short pulses, in other words, we will consider the limit $t_1\rightarrow T$, which resembles the delta driving case. Thus, in the delta driving layout, we turn on the deformation field given by Eq. (\ref{zdef}) at times $t=mT$ , while for $t\neq mT$ the deformation field is turned off, here $m$ is an integer number. A graphic representation of this driving layout is shown in Fig. \ref{layout}. }Within this limit ($t_1\rightarrow T$), the time-dependent Hamiltonian (\ref{shortpulse}) takes the following form,
\begin{equation}
H(k_x,t)=H_0(k_x)+\sum_{m}\left[H_1(k_x)-H_0(k_x)\right]\delta(t/T-m),
\end{equation}
with the Hamiltonians $H_0(k_x)$ and $H_1(k_x)$ given by,
\begin{equation}
H_0(k_x)=\sum^{N-1}_{j=1}\gamma_{0} \left[a_{2j+1}^{\dag} b_{2j}+c(k_x)a_{2j-1}^{\dag} b_{2j}\right]+\mathrm{h.c.},
\label{H0}
\end{equation}
and
\begin{equation}
H_1(k_x)=\sum^{N-1}_{j=1} \left[\gamma_{2j}\,a_{2j+1}^{\dag} b_{2j}+c(k_x)\, \gamma_{2j-1} a_{2j-1}^{\dag} b_{2j}\right]+\mathrm{h.c.}
\label{H1}
\end{equation}
{ Before entering into the details of our model, let us briefly discuss the effect of considering a sinusoidal time perturbation instead of a Dirac delta protocol. The Dirac delta driving is useful because calculations are greatly simplified and because analytical results can be obtained. One can consider a more realistic time perturbation but the system must be treated numerically. Consider for example a cosine-like driving, then the quasienergies of the system are given by the eigenvalues of the so-called Floquet Hamiltonian\cite{rudner2013anomalous}, which is a block diagonal matrix (for our case, each block is $N\times N$ matrix with $N$ being the number of atoms per unit cell). By truncating such Hamiltonian (this is, by considering only the first three blocks of such Hamiltonian), one can obtain numerically the quasienergies. By using this kind of driving as we have proven in a previous work\cite{roman2017topological} for a model quite similar to the one studied here, the secular gaps are reduced in size when compared with the delta driving. Additionally, the flat bands become dispersive edge states\cite{roman2017topological}. Summarizing, the emergence of highly localized edge states is not modified if a more realistic driving layout is considered.}

To study the time evolution of our system, we define the unitary one-period time evolution operator, $U(k_x,T)$, in the usual form,
\begin{equation}
U(k_x,T)\ket{\psi_{k_x}(t)}=\ket{\psi_{k_x}(t+T)}
\end{equation}
where $\ket{\psi_{k_x}(t)}$ is the system wave function for a given $k_x$. The main advantage of using a delta kicking is that the time evolution operator is  easy to find. For this case, we have,
\begin{equation}
\begin{split}
U(\tau)&=\mathcal{T}\exp{\left[-i\int_{0}^TH(k_x,t)\,dt/\hbar\right]}\\
 &=\exp{\left[-i\tau (H_1(k_x)-H_0(k_x))\right]}\exp{\left[-i\tau H_0(k_x)\right]},
\end{split}
\label{uop}
\end{equation}
here $\mathcal{T}$ denotes the time ordering operator and $\tau=T/\hbar$. In general Hamiltonians $H_1$ and $H_0$ do not commute, therefore, it is a common practice to study the eigenvalue spectrum of the matrix representation of Eq. (\ref{uop}) via an effective Hamiltonian defined as 
\begin{equation}
U(k_x,\tau)=\exp{\left[-i\tau H_\text{eff}(k_x)\right]}.
\end{equation}
Then, the eigenvalues of the time-evolution operator, which we denote by $\tau\omega$, are the eigenvalues of the effective Hamiltonian, $\tau H_\text{eff}(k_x)$. Since $\tau\omega$ are just defined up to integer multiples of $2\pi$, they are called the quasienergies of the system.

Once that the time-dependence have been introduced to our model, we have four free parameters, three owing to the deformation field ($\lambda$, $\sigma$, and $\phi$) and one due to the driving layout ($\tau$). One can study the quasienergy spectrum for a wide range of parameters, however just a few set of parameters allows us to do analytical calculations. Among them, one can mention the case $\sigma=1/3$ and $\phi=0$ for which the system becomes periodic both along the $x$-direction and the $y$-direction. This is due to the fact that the hopping parameters, for this particular case, just take two different values, namely,
\begin{equation}
\begin{split}
\gamma_{j}&=\left(1+\alpha-\frac{\alpha}{\sqrt{\frac{\pi ^2 \lambda ^2}{3}+1}}\right)\exp{\left[\beta  \left(1-\sqrt{\xi_j\lambda^2+1}\right)\right]},
\end{split}
\label{hoppings}
\end{equation}
where $\xi_j=1/4$ for odd $j$ and $\xi_j=3/2$ otherwise.

It is noteworthy that for $\sigma=1/3$, our system is quite similar to the system studied in reference \cite{roman2017topological}, therein a periodically driven uniaxial strained zigzag graphene nanoribbon is studied. The main result of such paper is the emergence of topologically protected flat bands at both zero and $\pm\pi$ quasienergies. The emergence of these flat bands can be understood in terms of a kind of Weyl points that appear each time that the bands are inverted\cite{Winkler17}. Therefore, we expect our model to have topological flat bands and Weyl points. This conjecture is confirmed in the next section where the touching band points of the quasienergy spectrum are found.

%%%%%%%%%%%% %%%%%%%%%%%%
\section{Touching band points}
\label{TBP}
%%%%%%%%%%%%%%%%%%%%%%%%

Our system can be studied numerically for any combination of driving parameters. From an analytical point of view, only few cases are simply enough to carry on calculations.  In fact, for inconnmensurate $\sigma$, the problem is very complex since quasiperiodicity arises and requires the use of rational approximants and renormalization approaches \cite{Naumis1993,Naumis1999,NaumisThorpe1999,Satija2013}. Here we have chosen to present simple analytical cases and compare it with the numerical results. 
In particular, we will study the quasienergy touching band points for $\sigma=1/3$, $\phi=0$ and fixed values of $\lambda$ and $\tau$. 
For this case, the system becomes periodic along both the $x$- and $y$-directions if cyclic boundary conditions are used in the $y$ axis. Nanoribbons are thus studied by changing the boundary conditions. This allows to define the Fourier transformed version of Hamiltonians Eqs. (\ref{H0}) and (\ref{H1}),  
\begin{equation}
\begin{split}
H_0(\mathbf{k})&=h_0(\mathbf{k})\,\mathbf{\hat{h}_0}(\mathbf{k})\cdot\mathbf{\sigma}\\
H_1(\mathbf{k})&=h_1(\mathbf{k})\,\mathbf{\hat{h}_1}(\mathbf{k})\cdot\mathbf{\sigma}
\end{split}
\label{FH}
\end{equation}
by using a vector in reciprocal space $\mathbf{k}=(k_x,k_y)$. $\sigma_{i}$ ($i=x,y,z$) are the $2\times2$ Pauli matrices and $\mathbf{\hat{h_0}(\mathbf{k})}=\mathbf{h_0(\mathbf{k})}/|h_0(\mathbf{k})|$, $\mathbf{\hat{h_1}(\mathbf{k})}=\mathbf{h_1}(\mathbf{k})/|h_1(\mathbf{k})|$. Here, $h_{0}(\mathbf{k})$ and $h_{1}(\mathbf{k})$) denote the norm of $\mathbf{h_{0}(\mathbf{k})}$ and $\mathbf{h}_{1}(\mathbf{k})$ respectively. $\mathbf{h_0}(\mathbf{k})$ and $\mathbf{h_1}(\mathbf{k})$ have components which are defined in appendix \ref{AA}. The $\mathbf{k}$-dependent time evolution operator, Eq. (\ref{uop}), now takes the following form,
\begin{equation}
U(\mathbf{k},\tau)=\sum_{k_y}\mathcal{U}(\mathbf{k},\tau)\otimes\ket{k_y}\bra{k_y},
\end{equation}
where,
\begin{equation}
\begin{split}
\mathcal{U}(\mathbf{k},\tau)=&\exp{\left[-i\tau \delta H(\mathbf{k})\right]}\exp{\left[-i\tau H_0(\mathbf{k})\right]}.
\end{split}
\label{Ueff}
\end{equation}
and $\delta H(\mathbf{k})=H_1(\mathbf{k})-H_0(\mathbf{k})$. To obtain the quasienergy spectrum we use an effective Hamiltonian approach. Let us define the effective Hamiltonian as,
\begin{equation}
\mathcal{U}(\mathbf{k},\tau)=\exp{\left[-i\tau H_{\mathrm{eff}}(\mathbf{k})\right]}.
\label{FUeff}
\end{equation}
%
%%%%%
\begin{figure}
\includegraphics[scale=0.315]{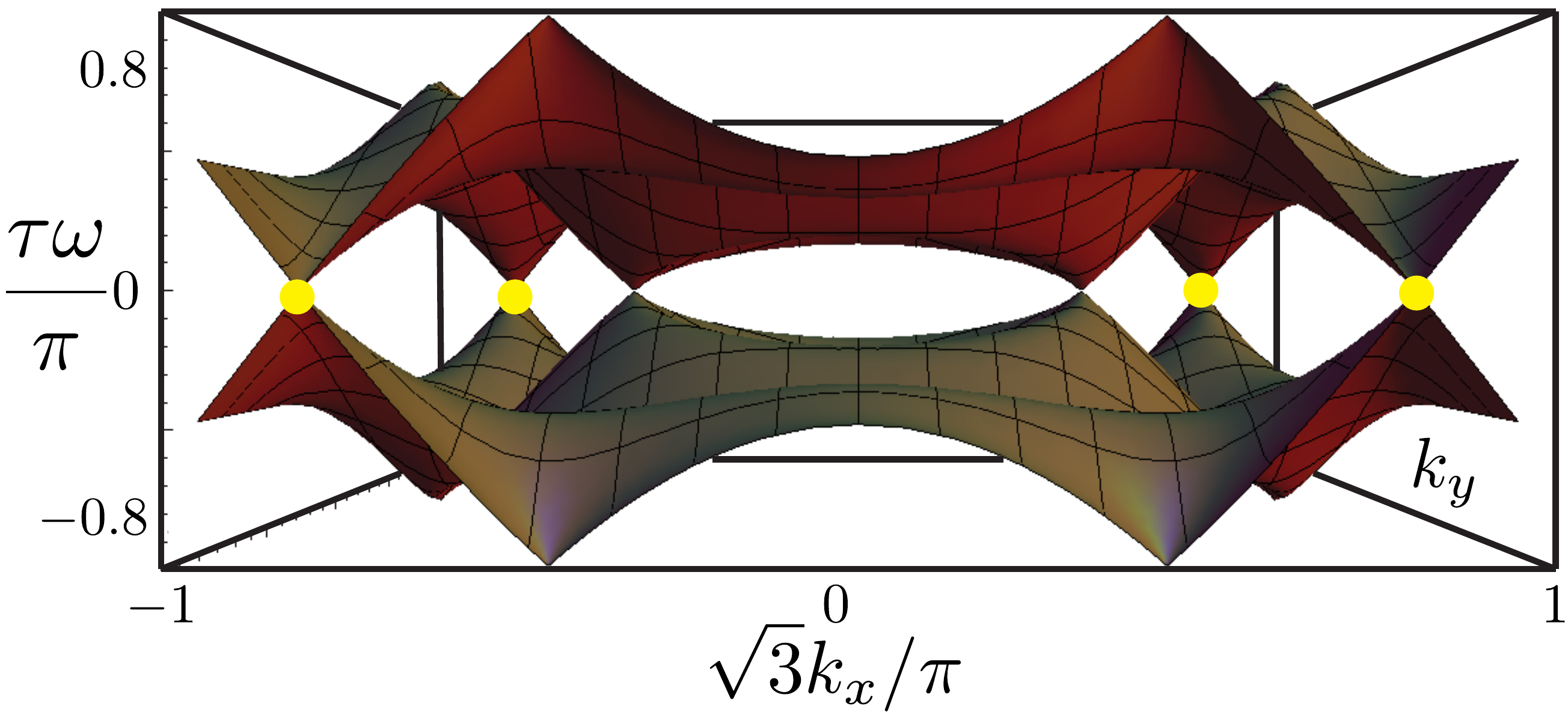}
\caption{(Color online). Quasienergy band structure as function of $\mathbf{k}$ for $\sigma=1/3$, $\phi=0$, $\lambda=0.5$, and $\tau=\pi$ obtained from the analytical expression Eq. (\ref{omega}). Note that besides the Dirac cones (which are shifted from their original positions due to the ripple field), indicated by yellow dots, others touching band points with linear dispersion around zero and $\pm\pi$ quasienergy emerge.}
\label{analyticQB}
\end{figure}
%%%%%
Since the Hamiltonians $H_0(\mathbf{k})$ and $H_1(\mathbf{k})$ are $2\times2$ matrices, it is possible to analytically obtain $H_{\mathrm{eff}}(\mathbf{k})$ using the addition rule of SU(2) (see appendix \ref{AA} for details). After some calculations and using Eqs. (\ref{FH}) and (\ref{Ueff}), one gets,
\begin{equation}
H_{\mathrm{eff}}(\mathbf{k})=\omega(\mathbf{k})\,\mathbf{\hat{h}_{\mathrm{eff}}(\mathbf{k})}\cdot\mathbf{\sigma},
\label{FHeff}
\end{equation}
and as before, $\mathbf{\sigma}$ is the Pauli vector. The quasienergies, $\tau\omega(\mathbf{k})$, are given by the following expression,
\begin{equation}
\begin{split}
&\cos{\left[\tau\omega(\mathbf{k})\right]}=\cos{[\tau\, \delta h(\mathbf{k})]}\cos{[\tau h_0(\mathbf{k})]}\\
&-\mathbf{\hat{h}_{0}(\mathbf{k})}\cdot\mathbf{\hat{\delta h}(\mathbf{k})}\sin{[\tau\,\delta h(\mathbf{k})]}\sin{[\tau h_0(\mathbf{k})]},
\end{split}
\label{omega}
\end{equation}
where $\mathbf{\delta h}(\mathbf{k})=\mathbf{h}_1(\mathbf{k})-\mathbf{h}_0(\mathbf{k})$, { and $\mathbf{\hat{h}_{\mathrm{eff}}}(\mathbf{k})$ is given by,
\begin{equation}
\begin{split}
\mathbf{\hat{h}_{\mathrm{eff}}}(\mathbf{k})&=\frac{-1}{\sin{\left[\tau\omega(\mathbf{k})\right]}}\left[\hat{\mathbf{\delta h}(\mathbf{k})}\sin{[\tau\,\delta h(\mathbf{k})]}\cos{[\tau h_0(\mathbf{k})]}\right]+\\
&\frac{-1}{\sin{\left[\tau\omega(\mathbf{k})\right]}}\left[\hat{\mathbf{h}}_0(\mathbf{k})\sin{[\tau h_0(\mathbf{k})]}\cos{[\tau \delta h(\mathbf{k})]}\right]+\\
&\frac{-1}{\sin{\left[\tau\omega(\mathbf{k})\right]}}\left[\hat{\mathbf{\delta h}(\mathbf{k})}\times\hat{\mathbf{h}}_0(\mathbf{k})\sin{[\tau \delta h(\mathbf{k})]}\sin{(\tau h_0[\mathbf{k})]}\right].
\end{split}
\label{unitvecMain}
\end{equation}
}
Since we are looking for touching band points, it is useful to plot the quasienergy spectrum for some characteristic values of $\lambda$ and $\tau$. In Fig. \ref{analyticQB} we plot the quasienergy band structure for $\sigma=1/3$, $\phi=0$, $\lambda=0.5$, and $\tau=\pi$. Note that apart the Dirac cones (indicated by yellow dots in the figure), there are other touching band points at zero and $\pm\pi$ quasienergies. 

From Fig. \ref{analyticQB}, we can see that touching band points always emerge at zero or $\pm\pi$ quasienergy, then it follows that they can be obtained by imposing $\tau\omega(\mathbf{k^*})=n\pi$, where $n$ is an integer number and $\mathbf{k^*}=(k_x^*,k_y^*)$ are the special points at which this happens. By substituting $\mathbf{k}=\mathbf{k^*}$ in Eq. (\ref{omega}), the touching band points are given by the solutions of the following equation,
\begin{equation}
\begin{split}
&\pm1=\cos{[\tau\, \delta h(\mathbf{k^*})]}\cos{[\tau h_0(\mathbf{k^*})]}\\
&-\mathbf{\hat{h}_{0}(\mathbf{k^*})}\cdot\mathbf{\hat{\delta h}(\mathbf{k^*})}\sin{[\tau\,\delta h(\mathbf{k^*})]}\sin{[\tau h_0(\mathbf{k^*})]}.
\end{split}
\label{omegat}
\end{equation}
A careful analysis of Eq. (\ref{omegat}) shows two possible solutions depending on the value of the dot product $\mathbf{\hat{h}_{0}(\mathbf{k^*})}\cdot\mathbf{\hat{\delta h}(\mathbf{k^*})}$. In other words, there are two kinds of touching band points that we have labeled by type I and type II. For the type I, it is required that $\mathbf{\hat{h}_{0}(\mathbf{k^*})}\cdot\mathbf{\hat{\delta h}(\mathbf{k^*})}=\pm1$, which is equivalent to ask the commutator $[H_1(\mathbf{k^*}),H_0(\mathbf{k^*})]$ to vanish. For type II, it is necessary to impose two simultaneous restrictions, the first one is $\mathbf{\hat{h}_{0}(\mathbf{k^*})}\cdot\mathbf{\hat{\delta h}(\mathbf{k^*})}\neq\pm1$, whereas the second one is given by $\cos{[\tau\delta h(\mathbf{k^*})]}\cos{[\tau h_0(\mathbf{k^*})]}=\pm1$, this means that type II touching band points never occur for $k_y^*=0, \pm2\pi/3$. It what follows, we will study the necessary conditions for having these kinds of touching band points. After that, the topological phase diagram of the system is obtained.

\subsection{Type I}
Although this kind of touching band points have been studied in a previous work for a very particular case of hopping parameters\cite{roman2017topological}, here we obtain the touching band points for the general case of an effective linear chain with two different hopping parameters, say $\gamma_1$ and $\gamma_2$.  We start our analysis by noticing from Eq. (\ref{cdot}), that $\mathbf{\hat{h}}_0(\mathbf{k}^{*})\cdot\mathbf{\hat{\delta h}}(\mathbf{k}^{*})=\pm1$ is fulfilled for $k_y^{*}=0,\pm 2\pi/3$, needless to say that such values of $k_y$ give the edges of the quasienergy band structure along the $y$-direction, we stress out the fact that at the edges of the quasienergy band structure, Hamiltonians $H_0(\mathbf{k})$ and $H_1(\mathbf{k})$ commute. By substituting $k_y^{*}$ into Eq. (\ref{omega}), one gets,
\begin{equation}
\tau\omega_{\pm}(k_x)=\tau\gamma_2\pm2\tau\gamma_1\cos{\left(\sqrt{3}k_x/2\right)}
\label{edges}
\end{equation}
where the plus sign ($+$) stems for $k_y^{*}=0$, while the minus sign ($-$) stems for $k_y^{*}=\pm2\pi/3$. Now, in order to have touching band points, two band edges must touch each other. This occurs whenever $\tau\omega(k_x^*)=\pm n\pi$ ($n$ being an integer number). By using Eq. (\ref{edges}), we find that $k_x^*$ has two possible solutions given by,
\begin{equation}
\begin{split}
k_x^{*(+)}&=\pm\frac{2}{\sqrt{3}}\arccos{\left[\frac{ n\pi-\tau\gamma_2}{2\tau\gamma_1}\right]}\\
k_x^{*(-)}&=\pm\frac{2}{\sqrt{3}}\arccos{\left[\frac{ -n\pi+\tau\gamma_2}{2\tau\gamma_1}\right]}.
\end{split}
\label{kx}
\end{equation}
As before, $k_x^{*(+)}$ stems for $k_y^{*}=0$, while $k_x^{*(-)}$ stems for $k_y^{*}=\pm2\pi/3$. From the structure of Eq. (\ref{kx}), it is easy to see that touching band points always come in pairs, as is the case of Weyl and Dirac points. We have to mention that for $n=0$ and for odd $n$ there are two pairs of touching band points, however this is not the case for even $n$ ($n$ different from zero) for which just one pair of touching band points emerge. This can be understood by looking at Eq. (\ref{kx}). It is readily seen that for even $n$ both $k_x^{+}$ and $k_x^{-}$ are the same. On the other hand, the case $n=0$ ({\it i.e.} the time-independent touching band points) worths special attention, since in this case the touching band points correspond to Dirac cones shifted from their original position due to the deformation field\cite{Pereira09}. As is well known, the Dirac cones give rise to flat bands in the time-independent case when the nanoribbon is considered to be finite, this still true even in the presence of a time-dependent deformation field\cite{roman2017topological}. As will be seen later on, touching band points for $n\neq0$ also give rise to topologically protected flat bands.

It is useful to obtain the conditions to have touching band points, since this sheds light about the topological phase diagram of the system. Such information can be readily obtained by observing that in order to have real solutions for Eq. (\ref{kx}), the following condition must be satisfied,
\begin{equation}
\left|n\pi-\tau\gamma_2\right|\leq 2\tau\gamma_1.
\label{phased}
\end{equation}
In other words, there is a critical treshold of $\tau$, say $\tau_c$ for having touching band points. Such value depends upon the ripple's amplitude via $\gamma_1$ and $\gamma_2$ (see Eq. (\ref{hoppings})). The explicit form of $\tau_c$ can be obtained from the extremal limits of Eq. (\ref{phased}), one can prove that is given by,
\begin{equation}
\tau_{c}=\frac{\pi}{2\gamma_1+\gamma_2}
\label{tauc}
\end{equation}
It is important to say that each time that $\tau$ reaches an integer multiple of $\tau_c$, new touching band points will emerge, in other words, there will be new pairs of touching band points for $\tau=n\tau_c$, where $n$ is an integer number. Also observe that bands will touch each other at $\pm\pi$ quasienergy if $n$ is odd, whereas they will touch each other at zero quasienergy for even or vanishing $n$. From Eq. (\ref{phased}), we can construct the phase diagram of type I touching band points, however, this phase diagram will be incomplete since it will not contain the information of the type II touching band points. Therefore, we leave the construction of the phase diagram to be done after analyzing type II touching band points. 

To finish, let us confirm our results. In Fig. \ref{analyticQB} we used $\lambda=0.5$ and $\tau=\pi$, this is, we have $2\tau_{c}^+<\tau<3\tau_c^{+}$. Therefore, there must be six pairs of touching band points, three pairs at zero quasienergy (two for $n=0$ and one for $n=2$) and two pairs at $\pm\pi$ quasienergy ($n=1$). This is in completely agreement with Fig. \ref{analyticQB}.

\subsection{Type II}

Let us start by determining the location of this kind of touching band points. To do that, we set $\tau\delta h=n_1\pi$ and $\tau h_0=n_2\pi$ in Eq. (\ref{omega}), where $n_1$ and $n_2$ are integer numbers. After some algebraic manipulations, one obtains,
\begin{equation}
\begin{split}
&\cos{(\sqrt{3}k^{*}_x/2)}=\\
&\pm\sqrt{\frac{\frac{n_1^2\pi^2}{\tau^2}-(\gamma_2-1)^2+(\gamma_1-1)(\gamma_2-2)\left(1-\frac{n_2^2\pi^2}{\tau^2}\right)}{4(\gamma_1-1)(\gamma_1-\gamma_2)}},\\
&\cos{(3k^{*}_y/2)}=\frac{n_2^2\pi^2-4\cos^2{(\sqrt{3}k^{*}_x/2)-1}}{4\cos{(\sqrt{3}k^{*}_x/2)}}.
\end{split}
\label{typeII}
\end{equation}
Once again, we can obtain the conditions for having these kind of touching band points by noticing that to ensure having real solutions in Eq. (\ref{typeII}), the following conditions need to be held altogether,
\begin{equation}
\begin{split}
&0\leq\frac{\frac{n_1^2\pi^2}{\tau^2}-(\gamma_2-1)^2+(\gamma_1-1)(\gamma_2-2)\left(1-\frac{n_2^2\pi^2}{\tau^2}\right)}{4(\gamma_1-1)(\gamma_1-\gamma_2)}\leq 1
\\
  &\left|\frac{n_2^2\pi^2-4\cos^2{(\sqrt{3}k^{*}_x/2)-1}}{4\cos{(\sqrt{3}k^{*}_x/2)}} \right|\leq1.
\end{split}
\label{phasedII}
\end{equation}
%

%%%%
\begin{figure}
\includegraphics[scale=0.415]{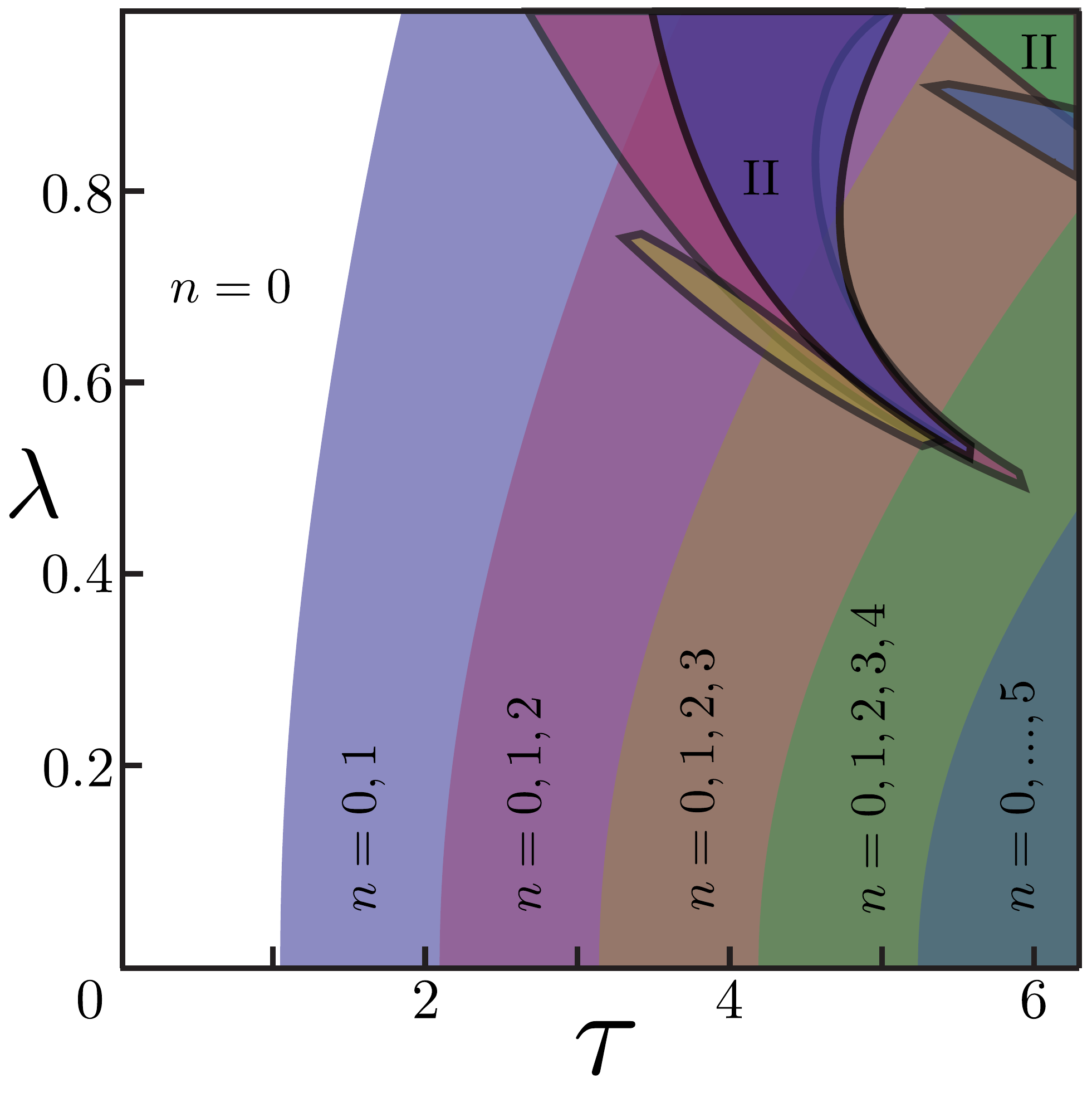}
\caption{(Color online). Phase diagram of the system for $\sigma=1/3$, $\phi=0$ obtained from the analytical expressions Eqs. (\ref{phased}) and (\ref{phasedII}). Two pairs of type I touching band points emerge for each value of $n$, each color corresponds to one value of $n$. { Regions that are not labeled by $n$ and that are surrounded by thick solid lines correspond to type II touching band points}. As can be seen, the phase diagram for type II touching band points is very complex and is located at high values of the ripple's amplitude. Therefore, their experimental observation may be hard.}
\label{phase}
\end{figure}
%%%%

It is worthwhile to mention that the band edges will touch each other at $\pm\pi$ quasienergy if $n_1$ is even and $n_2$ is odd or vice versa , whereas they will touch each other at zero quasienergy for either $n_1$ and $n_2$ even or odd. 

The conditions Eq. (\ref{phasedII}) add new phases to the phase diagram of the system. Such diagram will be built in the next section.

%%%%%%%%%
\subsection{Topological phase diagram}
%%%%%%%%%

In Fig. \ref{phase}, the phase diagram for type I and II touching band points is presented, built from the expressions for the critical values of $\tau$ obtained from Eq. (\ref{tauc}) and Eq. (\ref{phasedII}). Therein, type I touching band points are labeled by $n$ and each single value of $n$ gives rise to two pairs of this kind of touching band points, for instance, the region label by $n=0,1$ has four pairs of touching band points, two pairs corresponding to $n=0$ at zero quasienergy (Dirac cones, as was discussed above) and the others two pairs at $\pm\pi$ quasienergy corresponding to $n=1$. Note also that each value of $n$ corresponds to a well defined region in the phase diagram. When it concerns to type II touching band points things become more complicated since each pair of integers ($n_1$, $n_2$) results in very intricate regions on the phase diagram, as is clearly seen in Fig. \ref{phase} in the regions labeled by II. Additionally, for having type II touching band points high values of the ripple's amplitude are required, which make them difficult to be observed experimentally  since non-linear effects may appear before reaching this regimen. Finally, note that the fact that both kinds of touching band points always come in pairs suggests that they can give rise to topologically protected edge modes if the system is considered to be finite, in fact, this is the case as is proven below. 

\section{Edge states}
\label{edgestates}
%%

%%%%
\begin{figure}
\includegraphics[scale=0.38]{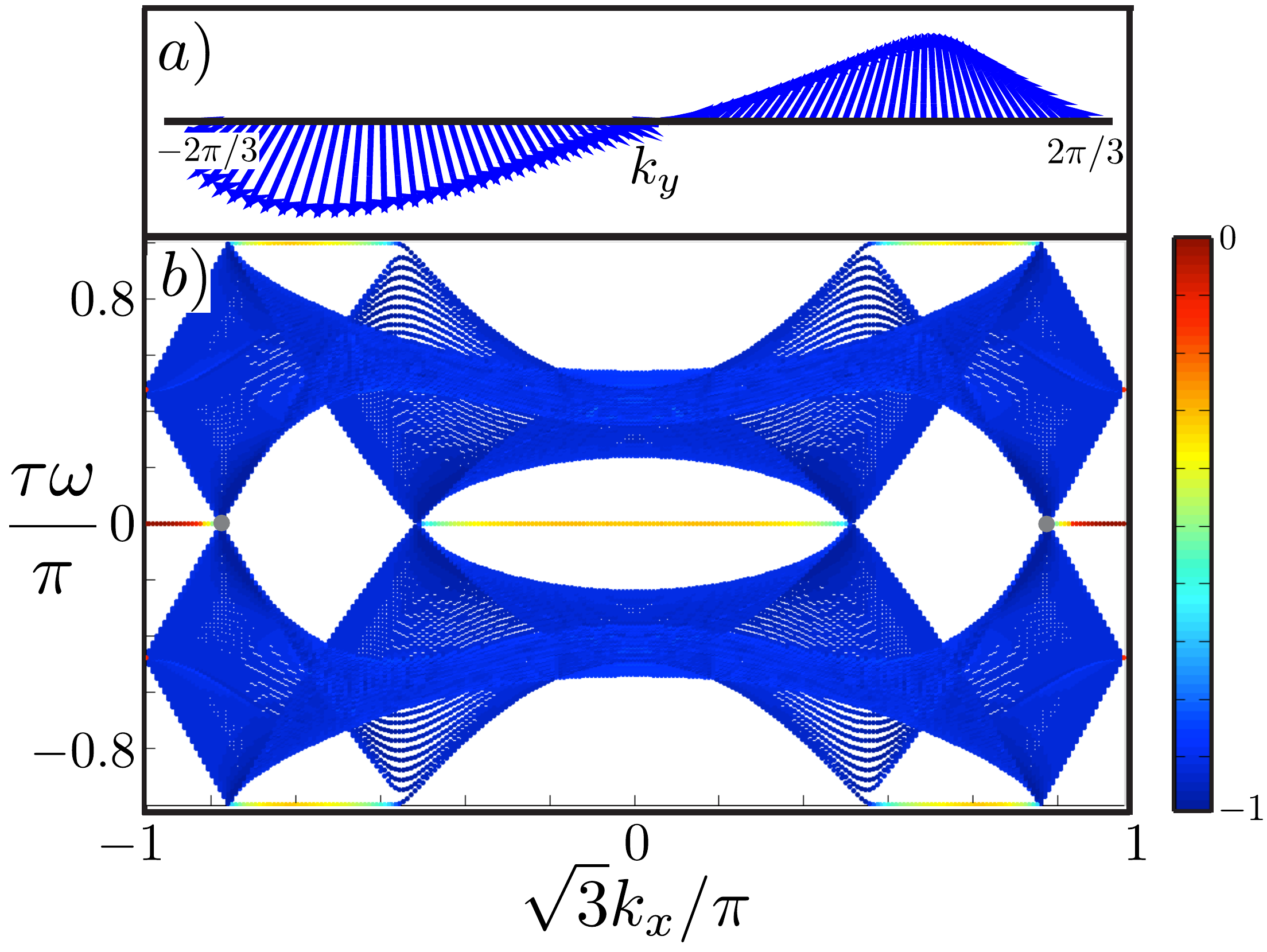}
\caption{(Color online). { In panel a) we present the winding of the vector $\hat{\mathbf{h}}_{\text{eff}}(\mathbf{k})$, obtained from the analytical expression Eq. (\ref{unitvecMain}) using $k_x=0.9\pi/\sqrt{3}$. The winding number of edge states that arise from the Dirac cones is one.} Panel b), quasienergy band structure obtained from the numerical diagonalization of Eq. (\ref{uop}) as a function of $k_x$ for $\sigma=1/3$, $\phi=0$, $\lambda=0.5$, $\tau=\pi$, and for a nanoribbon with $N=164$ atoms, also fixed boundary conditions were used. Note the excellent agreement between this plot and its analytical counterpart Fig. \ref{analyticQB}. In addition, observe the presence of flat bands at zero and $\pm\pi$ quasienergies, as predicted in the phase diagram Fig. \ref{phase} for type I touching band points with $n=0,1,2$. For $n=0$ we have Dirac cones (indicated by gray dots) shifted from their original positions due to the deformation field. The colors in the plot represent the logarithm of the inverse participation ratio, blue color corresponds to totally delocalized states, while red color stems for completely localized states. }
\label{qbs}
\end{figure}
%%%%

In this section we discuss the emergence and the topological properties of edge states in a finite zigzag graphene nanoribbon. In the previous section we found touching band points at which the edges of the quasienergy spectrum cross each other, which is a signature for edge states. In order to confirm if edge states emerge, we calculate the quasienergy spectrum for a finite system, to do that, a numerical diagonalization of the matrix representation of the time evolution operator Eq. (\ref{uop}), as a function of $k_x$, is done for fixed $\sigma$, $\phi$, $\lambda$ and $\tau$. We also study the localization properties of the wave functions of such states. To do that, we introduce the logarithm of the inverse participation ratio, which is defined as,
\begin{equation}
IPR(E)=\frac{\sum_{j=1}^{N}|\psi(j)|^4}{\ln{N}}.
\label{ipr}
\end{equation}
where $\psi(j)$ is the wave function at site $j$ for a given energy (or quasienergy) $E$. The IPR is a measure of the wave function localization\cite{Nosotros14}. The closer the IPR to zero the more localized the wave function is. Whereas for the IPR tending to $-1$, we have completely delocalized wave functions. Having said that, we proceed with the study of the edge states.

%%%%%%%
\subsection{Type I}
%%%%%%%

%%%%
\begin{figure}
\includegraphics[scale=0.42]{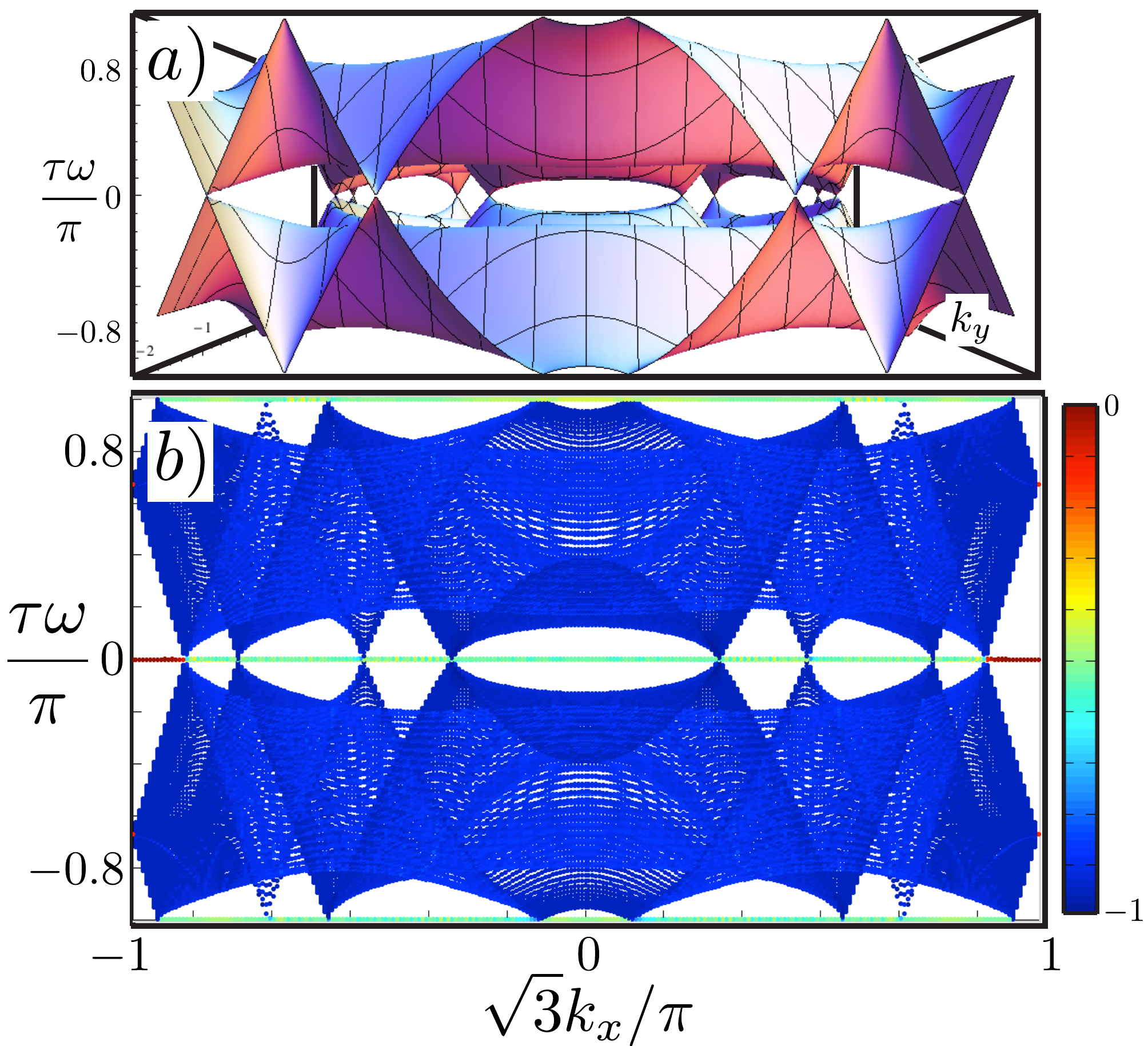}
\caption{{ (Color online). Panel a), quasienergy band structure obtained from the analytical expression (\ref{omega}) as a function of $k_x$ for $\sigma=1/3$, $\phi=0$, $\lambda=0.6$, and $\tau=6$. The parameters were chosen to be on a phase where only type I touching band points are observed. The maximum value of $n$ for these parameters is $4$ (see Eq. \ref{phasedII}). In panel b), we present the band structure of the system obtained from the numerical diagonalization of Eq. (\ref{uop}) for the same parameters used in panel a) but using fixed boundary conditions. The same color code as in Fig. \ref{qbs} was used. Note the emergence of flat bands that are less localized when compared with the ones observed in Fig. \ref{qbs}. The agreement between the numerical and analytical results is excellent.}}
\label{typeIbands}
\end{figure}
%%%%

Let us consider first the case of type I touching band points. We start by obtaining the quasienergy band structure as a function of $k_x$ via the numerical diagonalization of the matrix representation of Eq. (\ref{uop}). In Fig. \ref{qbs} we shown the resulting quasienergy band structure for $\sigma=1/3$, $\phi=0$, $\lambda=0.5$, $\tau=\pi$, $N=164$ atoms and obtained by using fixed boundary conditions. We used the same condition as in the analytically obtained plot in Fig. \ref{analyticQB} b). Note the excellent agreement between the numerical and the analytical results. 

{  In figure \ref{qbs} a) we also show the winding number of the effective Hamiltonian, which is basically the winding number of the unit vector defined in Eq. (\ref{unitvecMain}) for $k_x=0.9\pi/\sqrt{3}$, for a phase with  flat bands joining two inequivalent Dirac cones. As can be seen, the winding number is one, as expected from the topological properties of a finite ZGN.}

{ The main difference between Fig. \ref{qbs} and Fig. \ref{analyticQB} (apart from the fact that Fig. \ref{analyticQB} is a three dimensional plot and Fig. \ref{qbs} is the projected band structure as a function of $k_x$) is that, for a finite nanoribbon, highly localized edge modes are clearly seen in Fig. \ref{qbs}. In addition, we can see more touching band points in Fig. \ref{analyticQB} than in Fig. \ref{qbs} since the former is a three-dimensional plot in perspective (we have plotted the front view of the band structure), whereas the latter is a projection of the full band structure. For example, instead of seeing four Dirac cones in Fig. \ref{qbs}, as happens in Fig. \ref{analyticQB}, we  just see two Dirac cones because the projection superposes each pair. Something similar happens with the other touching band points.} The colors used in Fig. \ref{qbs} represent the logarithm of the inverse participation ratio (IPR, as defined in Eq. (\ref{ipr})), blue colors correspond to totally delocalized states and red color represents highly localized wave functions. Also observe how flat bands join two inequivalent touching band points, which suggests that inequivalent touching band points at the same quasienergy have opposite Berry phase. In fact, this is the case for $n=0$, which corresponds to Dirac cones, labeled by gray dots in Fig. \ref{qbs}. This also happens for $n\neq0$. 
{ Before studying the Berry phase of the touching band points and for the sake of clarity, in Fig. \ref{typeIbands} we present the analytical and the numerical band structure of our system for $\sigma=1/3$, $\phi=0$, $\tau=6$, and $\lambda=0.6$. These parameters were chosen in such a way that only type I touching band points appear. In panel Fig. \ref{typeIbands} a) we can observe many touching band points at zero and $\pm\pi$ quasienergies. Each pair produces flat bands as seen in panel b) of the same figure. It is important to note that the flat bands become more extended as the driving period is increased.}

To confirm the previous conjecture about the topological nature of the touching band points, we explicitly evaluate the Berry phase for type I touching band points. To do that, we start by noticing that near the touching band points the quasienergy spectrum is well described by the one-period time evolution operator, Eq. (\ref{Ueff}), expanded up to second order in powers of $\tau$. By using the Baker-Campbell-Hausdorff formula in Eq. (\ref{Ueff}), one gets,
\begin{equation}
\mathcal{U}(\mathbf{k},\tau) \approx \exp{\left\{-i\tau H_1(\mathbf{k})+\tau^2 [H_1(\mathbf{k}),H_0(\mathbf{k})]/2\right\}}.
\label{hserie}
\end{equation}
Since we are just interested in what happens in the neighborhood of touching band points, we expand Eq. (\ref{hserie}) around $(k_x^{*},k_y^{*})$. It is straightforward to show that Eq. (\ref{hserie}) can be written as 
\begin{equation}
\mathcal{U}(q_x,q_y,\tau)\approx \exp{\left[-i h_{T}\,\mathbf{\hat{h}}_{T}\cdot\mathbf{\sigma}\right]},
\label{approxU}
\end{equation}
where $q_x=k_x-k_x^{*}$, $q_y=k_y-k_y^{*}$, $h_{T}=|\mathbf{h}_{T}|$, and the vector $\mathbf{h}_{T}$ is given by,
\begin{equation}
\mathbf{h}_{T}=A(\lambda,\tau)\,q_x\hat{e}_x+B(\lambda,\tau)\,q_y\hat{e}_y+C(\lambda,\tau)\,q_y\hat{e}_z,
\label{topHam}
\end{equation}
with
\begin{equation}
\begin{split}
A(\gamma_1,\gamma_2,\tau)&=n\pi+\tau\gamma_1\sqrt{3+3\left(\frac{n\pi-\tau\gamma_2}{2\tau\gamma_1}\right)^2}\\
B(\gamma_1,\gamma_2,\tau)&=3\tau\gamma_2/2\\
C(\gamma_1,\gamma_2,\tau)&=\frac{3\tau}{4\gamma_1}(\gamma_1-\gamma_2)(n\pi-\tau\gamma_2).
\end{split}
\end{equation}
Finally, the Berry phase can be readily obtained from the effective Hamiltonian $h_{T}\,\mathbf{\hat{h}}_{T}\cdot\mathbf{\sigma}$. As prove in appendix \ref{BB}, the Berry phase, $\gamma_C$, is non-vanishing for touching band points at $k_x^{+}$, in fact, its value is $\gamma_C=\pi$. For touching band points at $k_x^{-}$ the Berry phase takes the opposite value as for $k_x^{+}$, this is, we have $\gamma_C=-\pi$. Therefore flat bands joining two touching band points with opposite Berry phase will emerge. Needless to say that these touching band points are topologically protected, so  flat bands are topologically non-trivial.

%%%%
\subsection{Type II}
%%%%

%%%%%
\begin{figure}
\includegraphics[scale=0.324]{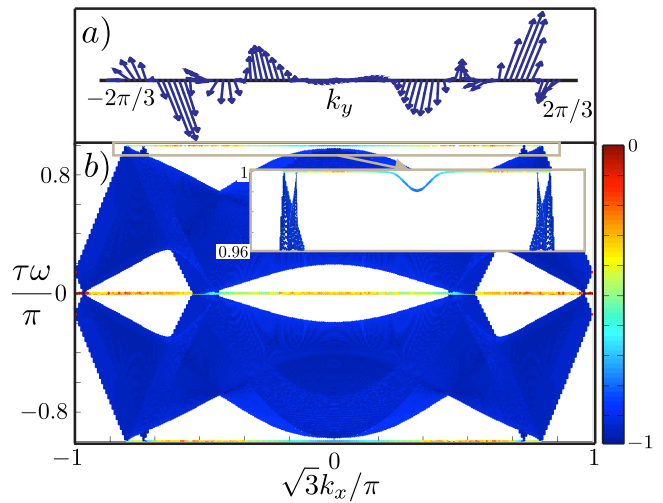}
\caption{(Color online). Panel a). Winding of the vector $\hat{\mathbf{h}}_{\text{eff}}(\mathbf{k})$ obtained from the analytical expression Eq. (\ref{unitvecMain}) for $k_x=0.9\pi/\sqrt{3}$, $\sigma=1/3$, $\phi=0$, $\tau=5.46$, and $\lambda=1$. {  By fixing $k_x$, we are studying a 1D slice of the system. The topological properties of this 1D slice are given by the winding of the unit vector $\hat{\mathbf{h}}_{\text{eff}}(\mathbf{k})$. A non-zero winding number is a signature of non-trivial topological properties.} Note that the winding number for this particular case is { $6$}. Panel b). Quasienergy band structure obtained from the numerical diagonalization of Eq. (\ref{uop}) as a function of $k_x$ for $\sigma=1/3$, $\phi=0$, $\tau=5.46$, $\lambda=1$, and $N=164$ using fixed boundary conditions. The same color code as in Fig. \ref{qbs} was used. Observe that for type II touching band points flat bands are less localized when compared with type I.}
\label{IIqbs}
\end{figure}
%%%%%

%%%%
\begin{figure}
\includegraphics[scale=0.42]{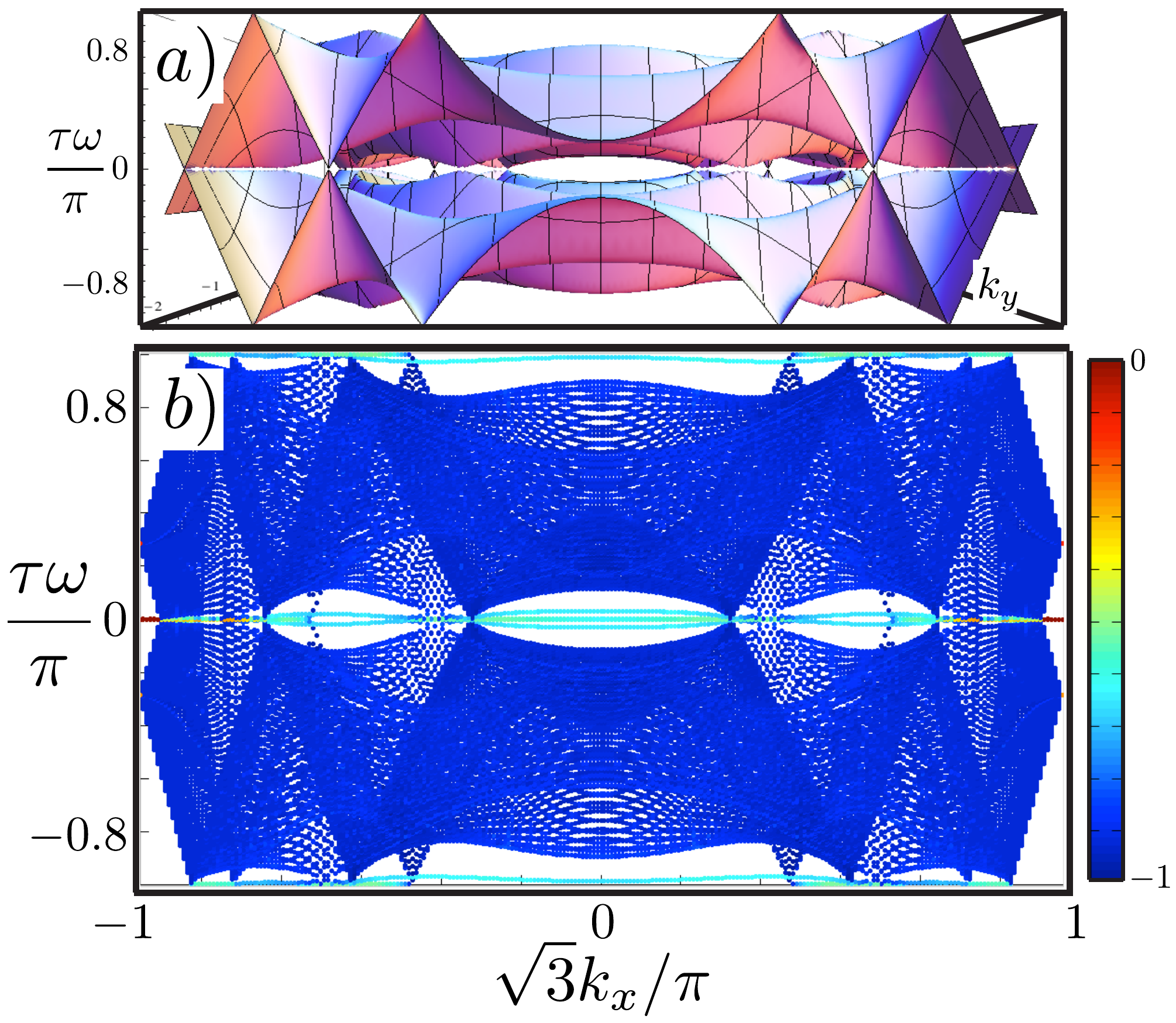}
\caption{{ (Color online). Panel a), quasienergy band structure obtained from the analytical expression (\ref{omega}) as a function of $k_x$ for $\sigma=1/3$, $\phi=0$, $\lambda=0.9$, and $\tau=7.5$. These parameters were chosen to have type I and type II touching band points. In panel b) we present the band structure of the system obtained from the numerical diagonalization of Eq. (\ref{uop}) for the same parameters as in panel a), but using fixed boundary conditions. The same color code as in Fig. \ref{qbs} was used. Due to the presence of type II touching band points, dispersive edge states appear. These dispersive edge states are almost extended. Once again the agreement between the numerical and analytical results is excellent.}}
\label{typeIIbands}
\end{figure}
%%%%

Now we analyze the edge states originated from type II touching band points. First, we obtain the quasienergy band structure from the numerical diagonalization of Eq. (\ref{uop}) for a set of parameters within one of the regions II of the diagram phase Fig. \ref{phase}. In Fig. \ref{IIqbs}, we show such band structure for $\sigma=1/3$, $\phi=0$, $\lambda=1$, $\tau=5.46$, and $N=164$, and obtained using fixed boundary conditions. Observe that in Fig. \ref{qbs} b) besides the type I touching band points there is one pair of type II touching band points. As in the case of type I touching band points, edge states emerge from type II touching band points, this edge states seem to be also flat bands. However, as the edge states approach to $k_x=0$, they are no longer flat bands but they become dispersive delocalized states, see the inset in Fig. \ref{IIqbs} b),  where a zoom around $\pm\pi$ quasienergy is shown. { To get further insight about the edge states that emerge from type II touching band points we plotted, in Fig. \ref{typeIIbands}, the analytical and numerical quasienergy band structure for $\sigma=1/3$, $\phi=0$, $\lambda=0.9$, and $\tau=7.5$. Observe that the agreement between the numerical [panel b)] and analytic [panel a)] results is quite good. As before, the edge states that appear in panel b) are dispersive and join two inequivalent touching band points. In addition, the edge states in Fig. \ref{typeIIbands} are less localized that the ones in Fig. \ref{typeIbands}.}

The fact that these edge states start and end at type II touching band points suggest that they have non-trivial topological properties. To study the topological properties of this kind of edge states we cannot proceed as we did with type I touching band points since type II touching band points do not correspond to points at where Hamiltonians (\ref{FH}) commute. Therefore, we analyze the topological properties of a one-dimensional slice of the system, in other words, we study our system for a fixed $k_x$. Once that we have fixed $k_x$, the topological properties can be obtained from the winding of the unit vector $\hat{\mathbf{h}}_{\text{eff}}$ that appears in the effective Hamiltonian Eq. (\ref{FHeff}), since a non-vanishing winding number is a signature of non-trivial topological properties. If $\hat{\mathbf{h}}_{\text{eff}}$, for fixed $k_x$, has a non-vanishing winding number around the origin, then the one dimensional slice has non-trivial topological properties and the whole two-dimensional (2D) system is topologically weak\cite{Yoshimura14,Ho14,WeakTop15}. In Fig. \ref{IIqbs} a) we show the winding of the unit vector $\hat{\mathbf{h}}_{\text{eff}}$ as a function of $k_y$ obtained from the analytical expression Eq. (\ref{unitvecMain}) for $k_x=0.9/\sqrt{3}$, $\sigma=1/3$, $\phi=0$, $\tau=5.46$, and $\lambda=1$. As clearly seen in the figure, the winding number is $6$, which means that our one-dimensional slice has non-trivial topological properties and that the whole 2D system is topologically weak.

%%%%%%%%%%%%%%%%%%
\section{Conclusions}
\label{conclusion}
%%%%%%%%%%%%%%%%%%

We have studied the case of a periodically driven rippled zigzag graphene nanoribbon. We obtained the quasienergy spectrum of the time-evolution operator. As a result, two types of touching band points were found for a special value of the corrugation wavelength ($\sigma=1/3$). Each type produces different  edge states. For type I edge states, we  found that the edge states are flat bands joining two inequivalent touching band points with opposite Berry phase, this was confirmed by analytical evaluation of the Berry phase. On the other hand, type II edge states were found to have a topological weak nature. This was done by a numerical calculation of the winding number of a one dimensional slice of the system, in other words, by looking at the topological properties of our system for a fixed $k_x$. Using this previous information, the phase diagram of the system was built. To finish, we stress out that the experimental realization of our model can be very challenging, however, there are some proposed experiments for similar situations\cite{koghee2012merging,zheng2014floquet,roman2017topological}. {  Experimentally is  possible to create a one-dimensional uniaxial ripple of graphene by growing it over a substrate\cite{bai2014creating}. Then the  driving can be achieved by time-periodically applying pressure to the whole system ({\it i.e.} to the graphene ribbon and  substrate). Time scales of femto seconds are needed to observe the phenomena discussed above,
a fact that requires the use of, for example, femto lasers of  Ti-Sapphire to induce deformations. As an alternative, optical lattices can be used since the hopping parameters can be tailored at will{ \cite{koghee2012merging,zheng2014floquet}}. 

Finally, it is important to remark that for observing the edge states studied here, the time driving layout does not need to be a delta driving. Even a cosine-like time perturbation can be used. However, for the case of a cosine-like time-perturbation, the effect could be hard to be observed since the secular gaps are usually smaller \cite{roman2017topological}.}

This work was supported by DGAPA-PAPIIT Project 102717. P. R.-T. acknowledges financial support from Consejo Nacional de Ciencia y Tecnolog\'ia (CONACYT) (M\'exico).

\vspace{1cm}

%%%%%%%%%%%%%%%%%%
\appendix
\section{}
\label{AA}
%%%%%%%%%%%%%%%%%%

In this appendix we analytically obtain the quasienergy spectrum for $\sigma=1/3$, $\phi=0$ and fixed values of $\lambda$ and $\tau$. As was mentioned in the main text, for $\sigma=1/3$, the system becomes periodic along both the $x$ and $y$ directions. As a result, we can Fourier transforming the Hamiltonians (\ref{H0}) and (\ref{H1}) taking advantage of such periodicity. By using the following Fourier transformations,
\begin{equation}
\begin{split}
a_j&=\frac{1}{\sqrt{N/2}}\sum_{k_y}e^{-i3k_y j/2}a_{k_y}\\
b_j&=\frac{1}{\sqrt{N/2}}\sum_{k_y}e^{-i3k_y j/2}b_{k_y},
\label{fourier}
\end{split}
\end{equation}
and after some algebraic manipulations, one gets the simplified Fourier transformed version of Hamiltonians Eq. (\ref{H0}) and (\ref{H1}),
\begin{equation}
\begin{split}
H_0(\mathbf{k})&=h_0(\mathbf{k})\,\mathbf{\hat{h}_0}(\mathbf{k})\cdot\mathbf{\sigma}\\
H_1(\mathbf{k})&=h_1(\mathbf{k})\,\mathbf{\hat{h}_1}(\mathbf{k})\cdot\mathbf{\sigma}
\end{split}
\label{hams}
\end{equation}
where $\mathbf{k}=(k_x,k_y)$, $\sigma_{i}$ ($i=x,y,z$) are the $2\times2$ Pauli matrices, $\mathbf{\hat{h_0}(\mathbf{k})}=\mathbf{h_0}(\mathbf{k})/|h_0(\mathbf{k})|$, $\mathbf{\hat{h_1}}(\mathbf{k})=\mathbf{h_1}/|h_1(\mathbf{k})|$ [$h_{0}(\mathbf{k})$ ($h_{1}(\mathbf{k})$) being the norm of $\mathbf{h_{0}}(\mathbf{k})$ ($\mathbf{h}_{1}(\mathbf{k})$)]. $\mathbf{h_0}(\mathbf{k})$ and $\mathbf{h_1}(\mathbf{k})$ have components given by
\begin{equation}
\begin{split}
h_0^{(x)}(\mathbf{k})&=2\cos{\left(\sqrt{3}k_x/2\right)}+\cos{\left(3k_y/2\right)}\\
h_0^{(y)}(\mathbf{k})&=\sin{\left(3k_y/2\right)},\\
h_1^{(x)}(\mathbf{k})&=2\gamma_1\cos{\left(\sqrt{3}k_x/2\right)}+\gamma_2\cos{\left(3k_y/2\right)}\\
h_1^{(y)}(\mathbf{k})&=\gamma_2\sin{\left(3k_y/2\right)},
\end{split}
\end{equation}
$\gamma_1$ and $\gamma_2$ have been defined in Eq. (\ref{hoppings}). By using Eq. (\ref{hams}), the time evolution operator Eq. (\ref{uop}) can be written as 
\begin{equation}
U(\mathbf{k},\tau)=\sum_{k_y}\mathcal{U}(\mathbf{k},\tau)\otimes\ket{k_y}\bra{k_y}.
\end{equation}
Here $\delta H(\mathbf{k})=H_1(\mathbf{k})-H_0(\mathbf{k})$, and 
\begin{equation}
\begin{split}
\mathcal{U}(\mathbf{k},\tau)=&\exp{\left[-i\tau \delta H(\mathbf{k})\right]}\exp{\left[-i\tau H_0(\mathbf{k})\right]}
\end{split}
\label{eqUeff}
\end{equation}
Even though, $H_1$ and $H_0$ generally do not commute, one can rewrite Eq. (\ref{eqUeff}) as follows,
\begin{equation}
\mathcal{U}(\mathbf{k},\tau)=\exp{\left[-i\tau H_{\mathrm{eff}}(\mathbf{k})\right]}.
\label{uopeff}
\end{equation}
where the effective Hamiltonian is given by
\begin{equation}
H_{\mathrm{eff}}(\mathbf{k})=\omega(\mathbf{k})\,\mathbf{\hat{h}_{\mathrm{eff}}}(\mathbf{k})\cdot\mathbf{\sigma},
\label{HZeff}
\end{equation}
the quasienergies $\tau\omega(\mathbf{k})$ are given by the next relation,
\begin{equation}
\begin{split}
&\cos{\left[\tau\omega(\mathbf{k})\right]}=\cos{[\tau\, \delta h(\mathbf{k})]}\cos{[\tau h_0(\mathbf{k})]}\\
&-\mathbf{\hat{h}_{0}}(\mathbf{k})\cdot\mathbf{\hat{\delta h}}(\mathbf{k})\sin{[\tau\,\delta h(\mathbf{k})]}\sin{[\tau h_0(\mathbf{k})]},
\end{split}
\label{omeganew}
\end{equation}
where $\mathbf{\delta h}(\mathbf{k})=\mathbf{h}_1(\mathbf{k})-\mathbf{h}_0(\mathbf{k})$, and
\begin{equation}
\begin{split}
&\mathbf{\hat{h}_{0}}(\mathbf{k})\cdot\mathbf{\hat{\delta h}}(\mathbf{k})=\frac{1}{h_0(\mathbf{k})\,\delta h(\mathbf{k})}\left[4(\gamma_1-1)\cos^2{\left(\frac{\sqrt{3}}{2}k_x\right)}\right]+\\
&\frac{1}{h_0(\mathbf{k})\delta h(\mathbf{k})}\left[2(\gamma_1+\gamma_2-2)\cos{\left(\frac{\sqrt{3}}{2}k_x\right)}\cos{\left(\frac{3k_y}{2}\right)}\right]+\\
&\frac{\gamma_2-1}{h_0(\mathbf{k})\delta h(\mathbf{k})}.
\end{split}
\label{cdot}
\end{equation}
Finally, the unit vector $\mathbf{\hat{h}}_{\text{eff}}(\mathbf{k})$ is given by
\begin{equation}
\begin{split}
\mathbf{\hat{h}_{\mathrm{eff}}}(\mathbf{k})&=\frac{-1}{\sin{\left[\tau\omega(\mathbf{k})\right]}}\left[\hat{\mathbf{\delta h}(\mathbf{k})}\sin{[\tau\,\delta h(\mathbf{k})]}\cos{[\tau h_0(\mathbf{k})]}\right]+\\
&\frac{-1}{\sin{\left[\tau\omega(\mathbf{k})\right]}}\left[\hat{\mathbf{h}}_0(\mathbf{k})\sin{[\tau h_0(\mathbf{k})]}\cos{[\tau \delta h(\mathbf{k})]}\right]+\\
&\frac{-1}{\sin{\left[\tau\omega(\mathbf{k})\right]}}\left[\hat{\mathbf{\delta h}(\mathbf{k})}\times\hat{\mathbf{h}}_0(\mathbf{k})\sin{[\tau \delta h(\mathbf{k})]}\sin{(\tau h_0[\mathbf{k})]}\right].
\end{split}
\label{unitvec}
\end{equation}
%

%%%%%%%%%%%%%
\section{}
\label{BB}
%%%%%%%%%%%%%

In this appendix, the explicit evaluation of the Berry phase for type I touching band points is done. The Berry phase is defined as
\begin{equation}
\gamma_{C}=\oint_{C}\mathbf{A}\cdot\,d\mathbf{k}
\label{Berry}
\end{equation}
where $\mathbf{A}=-i\bra{\psi_k}\mathbf{\nabla}_{k}\ket{\psi_k}$ is the so-called Berry connection (a gauge invariant quantity), and $\nabla_{k}=(\partial_{k_x},\partial_{k_y})$ is the gradient operator in the momentum space. Since we are interested in what happens in the neighborhood of type touching band points, it is enough to calculate the Berry phase of $\mathbf{\hat{h}}_{T}\cdot\mathbf{\sigma}$, which is the effective Hamiltonian in the neighborhood of type I touching band points and that is defined in Eq. (\ref{topHam}).

To obtain the Berry phase, we first need to calculate the eigenvectors of Hamiltonian Eq. (\ref{topHam}), it can be proven that such eigenvectors are given by the following spinors,
\begin{equation}
\begin{split}
\ket{\psi_{q'}^{\uparrow}}&= \frac{1}{\sqrt{2}}\left( \begin{array}{ccc}
\sqrt{1+\frac{C\,q{'}_y}{B\,h_T}} \\
e^{i\xi\alpha_{q{'}}}\sqrt{1-\frac{C\,q{'}_y}{B\,h_T}} \end{array} \right)\\
\ket{\psi_{q{'}}^{\downarrow}}&= -\frac{1}{\sqrt{2}}\left( \begin{array}{ccc}
e^{-i\xi\alpha_{q{'}}}\sqrt{1-\frac{C\,q{'}_y}{B\,h_T}} \\
 -\sqrt{1+\frac{C\,q{'}_y}{B\,h_T}}\end{array} \right)
\end{split}
\end{equation}
where
\begin{equation}
\begin{split}
q{'}_x&=q_x/A\\
q{'}_y&=q_y/B.
\end{split}
\end{equation}
and $\alpha_{q^{\prime}}$ is given by,
\begin{equation}
\alpha_{q{'}}=\tan^{-1}{\left(\frac{q{'}_y}{q{'}_x}\right)}.
\end{equation}
$\xi$ can take the values $\xi=+1$ which corresponds to $+k_x^{*(+)}$ and $\xi=-1$ to $-k_x^{*(+)}$. Now, the Berry connection can be calculated using such spinors, for simplicity we set $\xi=1$, however the result does not depend upon $\xi$. After some calculations, one obtains that the Berry connection is,
\begin{equation}
\mathbf{A}=\frac{1}{2}\left(1-\frac{C}{B\,h_{T}}q{'}_y\right)\nabla_{q{'}}\alpha_{q{'}},
\end{equation}
where 
\begin{equation}
\nabla_{q{'}}\alpha_{q{'}}=\frac{-q{'}_y\,\mathbf{\hat{e}}_x+q{'}_x\,\mathbf{\hat{e}}_y}{(q{'}_x)^2+(q{'}_y)^2}.
\end{equation}
Finally, we calculate the Berry phase along a circumference centered at $q{'}_x=q{'}_y=0$. By using polar coordinates, defined as, $q{'}_x=q{'}\cos{\theta}$ and $q{'}_y=q{'}\sin{\theta}$ where $(q{'})^2=(q{'}_x)^2+(q{'}_y)^2$, the Berry connection is readily obtained,
\begin{equation}
\begin{split}
\gamma_C&=\int_{0}^{2\pi}\,\mathbf{A}\cdot d\mathbf{q{'}}\\
&=\frac{1}{2}\int_{0}^{2\pi}\left(1-\frac{\frac{C}{B}\sin{\theta}}{\sqrt{1+\frac{C^2}{B^2}\sin^2{\theta}}}\right)d\theta=\pi.
\end{split}
\end{equation}

A similar calculation can be done for $k_x^{*(-)}$, which gives $\gamma_{C}=-\pi$.

\bibliography{biblioArmChairGraphene}{}

\begin{thebibliography}{89}%
\makeatletter
\providecommand \@ifxundefined [1]{%
 \@ifx{#1\undefined}
}%
\providecommand \@ifnum [1]{%
 \ifnum #1\expandafter \@firstoftwo
 \else \expandafter \@secondoftwo
 \fi
}%
\providecommand \@ifx [1]{%
 \ifx #1\expandafter \@firstoftwo
 \else \expandafter \@secondoftwo
 \fi
}%
\providecommand \natexlab [1]{#1}%
\providecommand \enquote  [1]{``#1''}%
\providecommand \bibnamefont  [1]{#1}%
\providecommand \bibfnamefont [1]{#1}%
\providecommand \citenamefont [1]{#1}%
\providecommand \href@noop [0]{\@secondoftwo}%
\providecommand \href [0]{\begingroup \@sanitize@url \@href}%
\providecommand \@href[1]{\@@startlink{#1}\@@href}%
\providecommand \@@href[1]{\endgroup#1\@@endlink}%
\providecommand \@sanitize@url [0]{\catcode `\\12\catcode `\$12\catcode
  `\&12\catcode `\#12\catcode `\^12\catcode `\_12\catcode `\%12\relax}%
\providecommand \@@startlink[1]{}%
\providecommand \@@endlink[0]{}%
\providecommand \url  [0]{\begingroup\@sanitize@url \@url }%
\providecommand \@url [1]{\endgroup\@href {#1}{\urlprefix }}%
\providecommand \urlprefix  [0]{URL }%
\providecommand \Eprint [0]{\href }%
\providecommand \doibase [0]{http://dx.doi.org/}%
\providecommand \selectlanguage [0]{\@gobble}%
\providecommand \bibinfo  [0]{\@secondoftwo}%
\providecommand \bibfield  [0]{\@secondoftwo}%
\providecommand \translation [1]{[#1]}%
\providecommand \BibitemOpen [0]{}%
\providecommand \bibitemStop [0]{}%
\providecommand \bibitemNoStop [0]{.\EOS\space}%
\providecommand \EOS [0]{\spacefactor3000\relax}%
\providecommand \BibitemShut  [1]{\csname bibitem#1\endcsname}%
\let\auto@bib@innerbib\@empty
%</preamble>
\bibitem [{\citenamefont {Katsnelson}(2007)}]{katsnelson07}%
  \BibitemOpen
  \bibfield  {author} {\bibinfo {author} {\bibfnamefont {M.~I.}\ \bibnamefont
  {Katsnelson}},\ }\href {\doibase
  http://dx.doi.org/10.1016/S1369-7021(06)71788-6} {\bibfield  {journal}
  {\bibinfo  {journal} {Materials Today}\ }\textbf {\bibinfo {volume} {10}},\
  \bibinfo {pages} {20 } (\bibinfo {year} {2007})}\BibitemShut {NoStop}%
\bibitem [{\citenamefont {Allen}\ \emph {et~al.}(2010)\citenamefont {Allen},
  \citenamefont {Tung},\ and\ \citenamefont {Kaner}}]{Review10}%
  \BibitemOpen
  \bibfield  {author} {\bibinfo {author} {\bibfnamefont {M.~J.}\ \bibnamefont
  {Allen}}, \bibinfo {author} {\bibfnamefont {V.~C.}\ \bibnamefont {Tung}}, \
  and\ \bibinfo {author} {\bibfnamefont {R.~B.}\ \bibnamefont {Kaner}},\ }\href
  {\doibase 10.1021/cr900070d} {\bibfield  {journal} {\bibinfo  {journal}
  {Chemical Reviews}\ }\textbf {\bibinfo {volume} {110}},\ \bibinfo {pages}
  {132} (\bibinfo {year} {2010})},\ \bibinfo {note} {pMID: 19610631},\ \Eprint
  {http://arxiv.org/abs/http://dx.doi.org/10.1021/cr900070d}
  {http://dx.doi.org/10.1021/cr900070d} \BibitemShut {NoStop}%
\bibitem [{\citenamefont {Carrillo-Bastos}\ \emph {et~al.}(2014)\citenamefont
  {Carrillo-Bastos}, \citenamefont {Faria}, \citenamefont {Latg\'e},
  \citenamefont {Mireles},\ and\ \citenamefont {Sandler}}]{Carrillo14}%
  \BibitemOpen
  \bibfield  {author} {\bibinfo {author} {\bibfnamefont {R.}~\bibnamefont
  {Carrillo-Bastos}}, \bibinfo {author} {\bibfnamefont {D.}~\bibnamefont
  {Faria}}, \bibinfo {author} {\bibfnamefont {A.}~\bibnamefont {Latg\'e}},
  \bibinfo {author} {\bibfnamefont {F.}~\bibnamefont {Mireles}}, \ and\
  \bibinfo {author} {\bibfnamefont {N.}~\bibnamefont {Sandler}},\ }\href
  {\doibase 10.1103/PhysRevB.90.041411} {\bibfield  {journal} {\bibinfo
  {journal} {Phys. Rev. B}\ }\textbf {\bibinfo {volume} {90}},\ \bibinfo
  {pages} {041411} (\bibinfo {year} {2014})}\BibitemShut {NoStop}%
\bibitem [{\citenamefont {Naumis}\ and\ \citenamefont
  {Roman-Taboada}(2014)}]{Nosotros14}%
  \BibitemOpen
  \bibfield  {author} {\bibinfo {author} {\bibfnamefont {G.~G.}\ \bibnamefont
  {Naumis}}\ and\ \bibinfo {author} {\bibfnamefont {P.}~\bibnamefont
  {Roman-Taboada}},\ }\href {\doibase 10.1103/PhysRevB.89.241404} {\bibfield
  {journal} {\bibinfo  {journal} {Phys. Rev. B}\ }\textbf {\bibinfo {volume}
  {89}},\ \bibinfo {pages} {241404} (\bibinfo {year} {2014})}\BibitemShut
  {NoStop}%
\bibitem [{\citenamefont {Oliva-Leyva}\ and\ \citenamefont
  {Naumis}(2014)}]{Olivaany14}%
  \BibitemOpen
  \bibfield  {author} {\bibinfo {author} {\bibfnamefont {M.}~\bibnamefont
  {Oliva-Leyva}}\ and\ \bibinfo {author} {\bibfnamefont {G.~G.}\ \bibnamefont
  {Naumis}},\ }\href {http://stacks.iop.org/0953-8984/26/i=12/a=125302}
  {\bibfield  {journal} {\bibinfo  {journal} {Journal of Physics: Condensed
  Matter}\ }\textbf {\bibinfo {volume} {26}},\ \bibinfo {pages} {125302}
  (\bibinfo {year} {2014})}\BibitemShut {NoStop}%
\bibitem [{\citenamefont {Oliva-Leyva}\ and\ \citenamefont
  {Naumis}(2013)}]{Maurice}%
  \BibitemOpen
  \bibfield  {author} {\bibinfo {author} {\bibfnamefont {M.}~\bibnamefont
  {Oliva-Leyva}}\ and\ \bibinfo {author} {\bibfnamefont {G.~G.}\ \bibnamefont
  {Naumis}},\ }\href {\doibase 10.1103/PhysRevB.88.085430} {\bibfield
  {journal} {\bibinfo  {journal} {Phys. Rev. B}\ }\textbf {\bibinfo {volume}
  {88}},\ \bibinfo {pages} {085430} (\bibinfo {year} {2013})}\BibitemShut
  {NoStop}%
\bibitem [{\citenamefont {Wang}\ \emph {et~al.}(2015)\citenamefont {Wang},
  \citenamefont {Wang},\ and\ \citenamefont {Liu}}]{wang2015generalized}%
  \BibitemOpen
  \bibfield  {author} {\bibinfo {author} {\bibfnamefont {B.}~\bibnamefont
  {Wang}}, \bibinfo {author} {\bibfnamefont {Y.}~\bibnamefont {Wang}}, \ and\
  \bibinfo {author} {\bibfnamefont {Y.}~\bibnamefont {Liu}},\ }\href {\doibase
  10.1142/S1793604715300017} {\bibfield  {journal} {\bibinfo  {journal}
  {Functional Materials Letters}\ }\textbf {\bibinfo {volume} {08}},\ \bibinfo
  {pages} {1530001} (\bibinfo {year} {2015})}\BibitemShut {NoStop}%
\bibitem [{\citenamefont {Bahamon}\ \emph {et~al.}(2015)\citenamefont
  {Bahamon}, \citenamefont {Qi}, \citenamefont {Park}, \citenamefont
  {Pereira},\ and\ \citenamefont {Campbell}}]{bahamon2015conductance}%
  \BibitemOpen
  \bibfield  {author} {\bibinfo {author} {\bibfnamefont {D.~A.}\ \bibnamefont
  {Bahamon}}, \bibinfo {author} {\bibfnamefont {Z.}~\bibnamefont {Qi}},
  \bibinfo {author} {\bibfnamefont {H.~S.}\ \bibnamefont {Park}}, \bibinfo
  {author} {\bibfnamefont {V.~M.}\ \bibnamefont {Pereira}}, \ and\ \bibinfo
  {author} {\bibfnamefont {D.~K.}\ \bibnamefont {Campbell}},\ }\href {\doibase
  10.1039/C5NR03393D} {\bibfield  {journal} {\bibinfo  {journal} {Nanoscale}\
  }\textbf {\bibinfo {volume} {7}},\ \bibinfo {pages} {15300} (\bibinfo {year}
  {2015})}\BibitemShut {NoStop}%
\bibitem [{\citenamefont {Roman-Taboada}\ and\ \citenamefont
  {Naumis}(2015)}]{roman15}%
  \BibitemOpen
  \bibfield  {author} {\bibinfo {author} {\bibfnamefont {P.}~\bibnamefont
  {Roman-Taboada}}\ and\ \bibinfo {author} {\bibfnamefont {G.~G.}\ \bibnamefont
  {Naumis}},\ }\href {\doibase 10.1103/PhysRevB.92.035406} {\bibfield
  {journal} {\bibinfo  {journal} {Phys. Rev. B}\ }\textbf {\bibinfo {volume}
  {92}},\ \bibinfo {pages} {035406} (\bibinfo {year} {2015})}\BibitemShut
  {NoStop}%
\bibitem [{\citenamefont {Salary}\ \emph {et~al.}(2016)\citenamefont {Salary},
  \citenamefont {Inampudi}, \citenamefont {Zhang}, \citenamefont {Tadmor},\
  and\ \citenamefont {Mosallaei}}]{Salary16}%
  \BibitemOpen
  \bibfield  {author} {\bibinfo {author} {\bibfnamefont {M.~M.}\ \bibnamefont
  {Salary}}, \bibinfo {author} {\bibfnamefont {S.}~\bibnamefont {Inampudi}},
  \bibinfo {author} {\bibfnamefont {K.}~\bibnamefont {Zhang}}, \bibinfo
  {author} {\bibfnamefont {E.~B.}\ \bibnamefont {Tadmor}}, \ and\ \bibinfo
  {author} {\bibfnamefont {H.}~\bibnamefont {Mosallaei}},\ }\href {\doibase
  10.1103/PhysRevB.94.235403} {\bibfield  {journal} {\bibinfo  {journal} {Phys.
  Rev. B}\ }\textbf {\bibinfo {volume} {94}},\ \bibinfo {pages} {235403}
  (\bibinfo {year} {2016})}\BibitemShut {NoStop}%
\bibitem [{\citenamefont {Oliva-Leyva}\ and\ \citenamefont
  {Naumis}(2016)}]{EffOliva16}%
  \BibitemOpen
  \bibfield  {author} {\bibinfo {author} {\bibfnamefont {M.}~\bibnamefont
  {Oliva-Leyva}}\ and\ \bibinfo {author} {\bibfnamefont {G.~G.}\ \bibnamefont
  {Naumis}},\ }\href {\doibase 10.1103/PhysRevB.93.035439} {\bibfield
  {journal} {\bibinfo  {journal} {Phys. Rev. B}\ }\textbf {\bibinfo {volume}
  {93}},\ \bibinfo {pages} {035439} (\bibinfo {year} {2016})}\BibitemShut
  {NoStop}%
\bibitem [{\citenamefont {Amorim}\ \emph {et~al.}(2016)\citenamefont {Amorim},
  \citenamefont {Cortijo}, \citenamefont {de~Juan}, \citenamefont {Grushin},
  \citenamefont {Guinea}, \citenamefont {Gutiérrez-Rubio}, \citenamefont
  {Ochoa}, \citenamefont {Parente}, \citenamefont {Roldán}, \citenamefont
  {San-Jose}, \citenamefont {Schiefele}, \citenamefont {Sturla},\ and\
  \citenamefont {Vozmediano}}]{Amorim16}%
  \BibitemOpen
  \bibfield  {author} {\bibinfo {author} {\bibfnamefont {B.}~\bibnamefont
  {Amorim}}, \bibinfo {author} {\bibfnamefont {A.}~\bibnamefont {Cortijo}},
  \bibinfo {author} {\bibfnamefont {F.}~\bibnamefont {de~Juan}}, \bibinfo
  {author} {\bibfnamefont {A.}~\bibnamefont {Grushin}}, \bibinfo {author}
  {\bibfnamefont {F.}~\bibnamefont {Guinea}}, \bibinfo {author} {\bibfnamefont
  {A.}~\bibnamefont {Gutiérrez-Rubio}}, \bibinfo {author} {\bibfnamefont
  {H.}~\bibnamefont {Ochoa}}, \bibinfo {author} {\bibfnamefont
  {V.}~\bibnamefont {Parente}}, \bibinfo {author} {\bibfnamefont
  {R.}~\bibnamefont {Roldán}}, \bibinfo {author} {\bibfnamefont
  {P.}~\bibnamefont {San-Jose}}, \bibinfo {author} {\bibfnamefont
  {J.}~\bibnamefont {Schiefele}}, \bibinfo {author} {\bibfnamefont
  {M.}~\bibnamefont {Sturla}}, \ and\ \bibinfo {author} {\bibfnamefont
  {M.}~\bibnamefont {Vozmediano}},\ }\href {\doibase
  http://dx.doi.org/10.1016/j.physrep.2015.12.006} {\bibfield  {journal}
  {\bibinfo  {journal} {Physics Reports}\ }\textbf {\bibinfo {volume} {617}},\
  \bibinfo {pages} {1 } (\bibinfo {year} {2016})},\ \bibinfo {note} {novel
  effects of strains in graphene and other two dimensional
  materials}\BibitemShut {NoStop}%
\bibitem [{\citenamefont {Carrillo-Bastos}\ \emph {et~al.}(2016)\citenamefont
  {Carrillo-Bastos}, \citenamefont {Le\'on}, \citenamefont {Faria},
  \citenamefont {Latg\'e}, \citenamefont {Andrei},\ and\ \citenamefont
  {Sandler}}]{Carrillo16}%
  \BibitemOpen
  \bibfield  {author} {\bibinfo {author} {\bibfnamefont {R.}~\bibnamefont
  {Carrillo-Bastos}}, \bibinfo {author} {\bibfnamefont {C.}~\bibnamefont
  {Le\'on}}, \bibinfo {author} {\bibfnamefont {D.}~\bibnamefont {Faria}},
  \bibinfo {author} {\bibfnamefont {A.}~\bibnamefont {Latg\'e}}, \bibinfo
  {author} {\bibfnamefont {E.~Y.}\ \bibnamefont {Andrei}}, \ and\ \bibinfo
  {author} {\bibfnamefont {N.}~\bibnamefont {Sandler}},\ }\href {\doibase
  10.1103/PhysRevB.94.125422} {\bibfield  {journal} {\bibinfo  {journal} {Phys.
  Rev. B}\ }\textbf {\bibinfo {volume} {94}},\ \bibinfo {pages} {125422}
  (\bibinfo {year} {2016})}\BibitemShut {NoStop}%
\bibitem [{\citenamefont {Hern{\'a}ndez-Ortiz}\ \emph
  {et~al.}(2016)\citenamefont {Hern{\'a}ndez-Ortiz}, \citenamefont
  {Valenzuela}, \citenamefont {Raya},\ and\ \citenamefont
  {S{\'a}nchez-Madrigal}}]{hernandez2016light}%
  \BibitemOpen
  \bibfield  {author} {\bibinfo {author} {\bibfnamefont {S.}~\bibnamefont
  {Hern{\'a}ndez-Ortiz}}, \bibinfo {author} {\bibfnamefont {D.}~\bibnamefont
  {Valenzuela}}, \bibinfo {author} {\bibfnamefont {A.}~\bibnamefont {Raya}}, \
  and\ \bibinfo {author} {\bibfnamefont {S.}~\bibnamefont
  {S{\'a}nchez-Madrigal}},\ }\href@noop {} {\bibfield  {journal} {\bibinfo
  {journal} {International Journal of Modern Physics B}\ }\textbf {\bibinfo
  {volume} {30}},\ \bibinfo {pages} {1650084} (\bibinfo {year}
  {2016})}\BibitemShut {NoStop}%
\bibitem [{\citenamefont {Sattari}(2016)}]{sattari2016spin}%
  \BibitemOpen
  \bibfield  {author} {\bibinfo {author} {\bibfnamefont {F.}~\bibnamefont
  {Sattari}},\ }\href@noop {} {\bibfield  {journal} {\bibinfo  {journal}
  {Journal of Magnetism and Magnetic Materials}\ }\textbf {\bibinfo {volume}
  {414}},\ \bibinfo {pages} {19} (\bibinfo {year} {2016})}\BibitemShut
  {NoStop}%
\bibitem [{\citenamefont {Sattari}\ and\ \citenamefont
  {Mirershadi}(2016)}]{sattari2016effects}%
  \BibitemOpen
  \bibfield  {author} {\bibinfo {author} {\bibfnamefont {F.}~\bibnamefont
  {Sattari}}\ and\ \bibinfo {author} {\bibfnamefont {S.}~\bibnamefont
  {Mirershadi}},\ }\href@noop {} {\bibfield  {journal} {\bibinfo  {journal}
  {The European Physical Journal B}\ }\textbf {\bibinfo {volume} {89}},\
  \bibinfo {pages} {227} (\bibinfo {year} {2016})}\BibitemShut {NoStop}%
\bibitem [{\citenamefont {Stegmann}\ and\ \citenamefont
  {Szpak}(2016)}]{stegmann2016current}%
  \BibitemOpen
  \bibfield  {author} {\bibinfo {author} {\bibfnamefont {T.}~\bibnamefont
  {Stegmann}}\ and\ \bibinfo {author} {\bibfnamefont {N.}~\bibnamefont
  {Szpak}},\ }\href@noop {} {\bibfield  {journal} {\bibinfo  {journal} {New
  Journal of Physics}\ }\textbf {\bibinfo {volume} {18}},\ \bibinfo {pages}
  {053016} (\bibinfo {year} {2016})}\BibitemShut {NoStop}%
\bibitem [{\citenamefont {L{\'o}pez-Sancho}\ and\ \citenamefont
  {Brey}(2016)}]{lopez2016magnetic}%
  \BibitemOpen
  \bibfield  {author} {\bibinfo {author} {\bibfnamefont {M.~P.}\ \bibnamefont
  {L{\'o}pez-Sancho}}\ and\ \bibinfo {author} {\bibfnamefont {L.}~\bibnamefont
  {Brey}},\ }\href@noop {} {\bibfield  {journal} {\bibinfo  {journal} {Physical
  Review B}\ }\textbf {\bibinfo {volume} {94}},\ \bibinfo {pages} {165430}
  (\bibinfo {year} {2016})}\BibitemShut {NoStop}%
\bibitem [{\citenamefont {Settnes}\ \emph
  {et~al.}(2016{\natexlab{a}})\citenamefont {Settnes}, \citenamefont {Leconte},
  \citenamefont {Barrios-Vargas}, \citenamefont {Jauho},\ and\ \citenamefont
  {Roche}}]{mikkel2016quantum}%
  \BibitemOpen
  \bibfield  {author} {\bibinfo {author} {\bibfnamefont {M.}~\bibnamefont
  {Settnes}}, \bibinfo {author} {\bibfnamefont {N.}~\bibnamefont {Leconte}},
  \bibinfo {author} {\bibfnamefont {J.~E.}\ \bibnamefont {Barrios-Vargas}},
  \bibinfo {author} {\bibfnamefont {A.-P.}\ \bibnamefont {Jauho}}, \ and\
  \bibinfo {author} {\bibfnamefont {S.}~\bibnamefont {Roche}},\ }\href
  {http://stacks.iop.org/2053-1583/3/i=3/a=034005} {\bibfield  {journal}
  {\bibinfo  {journal} {2D Materials}\ }\textbf {\bibinfo {volume} {3}},\
  \bibinfo {pages} {034005} (\bibinfo {year} {2016}{\natexlab{a}})}\BibitemShut
  {NoStop}%
\bibitem [{\citenamefont {Si}\ \emph {et~al.}(2016)\citenamefont {Si},
  \citenamefont {Sun},\ and\ \citenamefont {Liu}}]{ChenSi16}%
  \BibitemOpen
  \bibfield  {author} {\bibinfo {author} {\bibfnamefont {C.}~\bibnamefont
  {Si}}, \bibinfo {author} {\bibfnamefont {Z.}~\bibnamefont {Sun}}, \ and\
  \bibinfo {author} {\bibfnamefont {F.}~\bibnamefont {Liu}},\ }\href {\doibase
  10.1039/C5NR07755A} {\bibfield  {journal} {\bibinfo  {journal} {Nanoscale}\
  }\textbf {\bibinfo {volume} {8}},\ \bibinfo {pages} {3207} (\bibinfo {year}
  {2016})}\BibitemShut {NoStop}%
\bibitem [{\citenamefont {Settnes}\ \emph
  {et~al.}(2016{\natexlab{b}})\citenamefont {Settnes}, \citenamefont {Power},\
  and\ \citenamefont {Jauho}}]{settnes2016pseudomagnetic}%
  \BibitemOpen
  \bibfield  {author} {\bibinfo {author} {\bibfnamefont {M.}~\bibnamefont
  {Settnes}}, \bibinfo {author} {\bibfnamefont {S.~R.}\ \bibnamefont {Power}},
  \ and\ \bibinfo {author} {\bibfnamefont {A.-P.}\ \bibnamefont {Jauho}},\
  }\href@noop {} {\bibfield  {journal} {\bibinfo  {journal} {Physical Review
  B}\ }\textbf {\bibinfo {volume} {93}},\ \bibinfo {pages} {035456} (\bibinfo
  {year} {2016}{\natexlab{b}})}\BibitemShut {NoStop}%
\bibitem [{\citenamefont {Naumis}\ \emph {et~al.}(2017)\citenamefont {Naumis},
  \citenamefont {Barraza-Lopez}, \citenamefont {Oliva-Leyva},\ and\
  \citenamefont {Terrones}}]{Review17}%
  \BibitemOpen
  \bibfield  {author} {\bibinfo {author} {\bibfnamefont {G.~G.}\ \bibnamefont
  {Naumis}}, \bibinfo {author} {\bibfnamefont {S.}~\bibnamefont
  {Barraza-Lopez}}, \bibinfo {author} {\bibfnamefont {M.}~\bibnamefont
  {Oliva-Leyva}}, \ and\ \bibinfo {author} {\bibfnamefont {H.}~\bibnamefont
  {Terrones}},\ }\href@noop {} {\bibfield  {journal} {\bibinfo  {journal}
  {Reports on Progress in Physics}\ } (\bibinfo {year} {2017})}\BibitemShut
  {NoStop}%
\bibitem [{\citenamefont {Diniz}\ \emph {et~al.}(2017)\citenamefont {Diniz},
  \citenamefont {Vernek},\ and\ \citenamefont
  {Souza}}]{diniz2017graphene-based}%
  \BibitemOpen
  \bibfield  {author} {\bibinfo {author} {\bibfnamefont {G.}~\bibnamefont
  {Diniz}}, \bibinfo {author} {\bibfnamefont {E.}~\bibnamefont {Vernek}}, \
  and\ \bibinfo {author} {\bibfnamefont {F.}~\bibnamefont {Souza}},\ }\href
  {\doibase http://dx.doi.org/10.1016/j.physe.2016.06.008} {\bibfield
  {journal} {\bibinfo  {journal} {Physica E: Low-dimensional Systems and
  Nanostructures}\ }\textbf {\bibinfo {volume} {85}},\ \bibinfo {pages} {264 }
  (\bibinfo {year} {2017})}\BibitemShut {NoStop}%
\bibitem [{\citenamefont {Akinwande}\ \emph {et~al.}(2017)\citenamefont
  {Akinwande}, \citenamefont {Brennan}, \citenamefont {Bunch}, \citenamefont
  {Egberts}, \citenamefont {Felts}, \citenamefont {Gao}, \citenamefont {Huang},
  \citenamefont {Kim}, \citenamefont {Li}, \citenamefont {Li}, \citenamefont
  {Liechti}, \citenamefont {Lu}, \citenamefont {Park}, \citenamefont {Reed},
  \citenamefont {Wang}, \citenamefont {Yakobson}, \citenamefont {Zhang},
  \citenamefont {Zhang}, \citenamefont {Zhou},\ and\ \citenamefont
  {Zhu}}]{Akinwande17}%
  \BibitemOpen
  \bibfield  {author} {\bibinfo {author} {\bibfnamefont {D.}~\bibnamefont
  {Akinwande}}, \bibinfo {author} {\bibfnamefont {C.~J.}\ \bibnamefont
  {Brennan}}, \bibinfo {author} {\bibfnamefont {J.~S.}\ \bibnamefont {Bunch}},
  \bibinfo {author} {\bibfnamefont {P.}~\bibnamefont {Egberts}}, \bibinfo
  {author} {\bibfnamefont {J.~R.}\ \bibnamefont {Felts}}, \bibinfo {author}
  {\bibfnamefont {H.}~\bibnamefont {Gao}}, \bibinfo {author} {\bibfnamefont
  {R.}~\bibnamefont {Huang}}, \bibinfo {author} {\bibfnamefont {J.-S.}\
  \bibnamefont {Kim}}, \bibinfo {author} {\bibfnamefont {T.}~\bibnamefont
  {Li}}, \bibinfo {author} {\bibfnamefont {Y.}~\bibnamefont {Li}}, \bibinfo
  {author} {\bibfnamefont {K.~M.}\ \bibnamefont {Liechti}}, \bibinfo {author}
  {\bibfnamefont {N.}~\bibnamefont {Lu}}, \bibinfo {author} {\bibfnamefont
  {H.~S.}\ \bibnamefont {Park}}, \bibinfo {author} {\bibfnamefont {E.~J.}\
  \bibnamefont {Reed}}, \bibinfo {author} {\bibfnamefont {P.}~\bibnamefont
  {Wang}}, \bibinfo {author} {\bibfnamefont {B.~I.}\ \bibnamefont {Yakobson}},
  \bibinfo {author} {\bibfnamefont {T.}~\bibnamefont {Zhang}}, \bibinfo
  {author} {\bibfnamefont {Y.-W.}\ \bibnamefont {Zhang}}, \bibinfo {author}
  {\bibfnamefont {Y.}~\bibnamefont {Zhou}}, \ and\ \bibinfo {author}
  {\bibfnamefont {Y.}~\bibnamefont {Zhu}},\ }\href {\doibase
  https://doi.org/10.1016/j.eml.2017.01.008} {\bibfield  {journal} {\bibinfo
  {journal} {Extreme Mechanics Letters}\ }\textbf {\bibinfo {volume} {13}},\
  \bibinfo {pages} {42 } (\bibinfo {year} {2017})}\BibitemShut {NoStop}%
\bibitem [{\citenamefont {Ghahari}\ \emph {et~al.}(2017)\citenamefont
  {Ghahari}, \citenamefont {Walkup}, \citenamefont {Guti{\'e}rrez},
  \citenamefont {Rodriguez-Nieva}, \citenamefont {Zhao}, \citenamefont
  {Wyrick}, \citenamefont {Natterer}, \citenamefont {Cullen}, \citenamefont
  {Watanabe}, \citenamefont {Taniguchi}, \citenamefont {Levitov}, \citenamefont
  {Zhitenev},\ and\ \citenamefont {Stroscio}}]{Ghahari17}%
  \BibitemOpen
  \bibfield  {author} {\bibinfo {author} {\bibfnamefont {F.}~\bibnamefont
  {Ghahari}}, \bibinfo {author} {\bibfnamefont {D.}~\bibnamefont {Walkup}},
  \bibinfo {author} {\bibfnamefont {C.}~\bibnamefont {Guti{\'e}rrez}}, \bibinfo
  {author} {\bibfnamefont {J.~F.}\ \bibnamefont {Rodriguez-Nieva}}, \bibinfo
  {author} {\bibfnamefont {Y.}~\bibnamefont {Zhao}}, \bibinfo {author}
  {\bibfnamefont {J.}~\bibnamefont {Wyrick}}, \bibinfo {author} {\bibfnamefont
  {F.~D.}\ \bibnamefont {Natterer}}, \bibinfo {author} {\bibfnamefont {W.~G.}\
  \bibnamefont {Cullen}}, \bibinfo {author} {\bibfnamefont {K.}~\bibnamefont
  {Watanabe}}, \bibinfo {author} {\bibfnamefont {T.}~\bibnamefont {Taniguchi}},
  \bibinfo {author} {\bibfnamefont {L.~S.}\ \bibnamefont {Levitov}}, \bibinfo
  {author} {\bibfnamefont {N.~B.}\ \bibnamefont {Zhitenev}}, \ and\ \bibinfo
  {author} {\bibfnamefont {J.~A.}\ \bibnamefont {Stroscio}},\ }\href {\doibase
  10.1126/science.aal0212} {\bibfield  {journal} {\bibinfo  {journal}
  {Science}\ }\textbf {\bibinfo {volume} {356}},\ \bibinfo {pages} {845}
  (\bibinfo {year} {2017})},\ \Eprint
  {http://arxiv.org/abs/http://science.sciencemag.org/content/356/6340/845.full.pdf}
  {http://science.sciencemag.org/content/356/6340/845.full.pdf} \BibitemShut
  {NoStop}%
\bibitem [{\citenamefont {Milovanovi{\'c}}\ \emph {et~al.}(2017)\citenamefont
  {Milovanovi{\'c}}, \citenamefont {Tadi{\'c}},\ and\ \citenamefont
  {Peeters}}]{milovanovic2017graphene}%
  \BibitemOpen
  \bibfield  {author} {\bibinfo {author} {\bibfnamefont {S.}~\bibnamefont
  {Milovanovi{\'c}}}, \bibinfo {author} {\bibfnamefont {M.}~\bibnamefont
  {Tadi{\'c}}}, \ and\ \bibinfo {author} {\bibfnamefont {F.}~\bibnamefont
  {Peeters}},\ }\href@noop {} {\bibfield  {journal} {\bibinfo  {journal}
  {Applied Physics Letters}\ }\textbf {\bibinfo {volume} {111}},\ \bibinfo
  {pages} {043101} (\bibinfo {year} {2017})}\BibitemShut {NoStop}%
\bibitem [{\citenamefont {Prabhakar}\ \emph {et~al.}(2017)\citenamefont
  {Prabhakar}, \citenamefont {Melnik},\ and\ \citenamefont
  {Bonilla}}]{prabhakar2017strain}%
  \BibitemOpen
  \bibfield  {author} {\bibinfo {author} {\bibfnamefont {S.}~\bibnamefont
  {Prabhakar}}, \bibinfo {author} {\bibfnamefont {R.}~\bibnamefont {Melnik}}, \
  and\ \bibinfo {author} {\bibfnamefont {L.}~\bibnamefont {Bonilla}},\ }\href
  {\doibase 10.1140/epjb/e2017-80038-3} {\bibfield  {journal} {\bibinfo
  {journal} {The European Physical Journal B}\ }\textbf {\bibinfo {volume}
  {90}},\ \bibinfo {pages} {92} (\bibinfo {year} {2017})}\BibitemShut {NoStop}%
\bibitem [{\citenamefont {Cariglia}\ \emph {et~al.}(2017)\citenamefont
  {Cariglia}, \citenamefont {Giamb\`o},\ and\ \citenamefont
  {Perali}}]{cariglia2017curvature}%
  \BibitemOpen
  \bibfield  {author} {\bibinfo {author} {\bibfnamefont {M.}~\bibnamefont
  {Cariglia}}, \bibinfo {author} {\bibfnamefont {R.}~\bibnamefont {Giamb\`o}},
  \ and\ \bibinfo {author} {\bibfnamefont {A.}~\bibnamefont {Perali}},\ }\href
  {\doibase 10.1103/PhysRevB.95.245426} {\bibfield  {journal} {\bibinfo
  {journal} {Phys. Rev. B}\ }\textbf {\bibinfo {volume} {95}},\ \bibinfo
  {pages} {245426} (\bibinfo {year} {2017})}\BibitemShut {NoStop}%
\bibitem [{\citenamefont {Bordag}\ \emph {et~al.}(2017)\citenamefont {Bordag},
  \citenamefont {Fialkovsky},\ and\ \citenamefont
  {Vassilevich}}]{bordag2017casimir}%
  \BibitemOpen
  \bibfield  {author} {\bibinfo {author} {\bibfnamefont {M.}~\bibnamefont
  {Bordag}}, \bibinfo {author} {\bibfnamefont {I.}~\bibnamefont {Fialkovsky}},
  \ and\ \bibinfo {author} {\bibfnamefont {D.}~\bibnamefont {Vassilevich}},\
  }\href@noop {} {\bibfield  {journal} {\bibinfo  {journal} {Physics Letters
  A}\ } (\bibinfo {year} {2017})}\BibitemShut {NoStop}%
\bibitem [{\citenamefont {Zhang}\ \emph {et~al.}(2017)\citenamefont {Zhang},
  \citenamefont {Pan},\ and\ \citenamefont {Wang}}]{zhang2017frequency}%
  \BibitemOpen
  \bibfield  {author} {\bibinfo {author} {\bibfnamefont {S.-J.}\ \bibnamefont
  {Zhang}}, \bibinfo {author} {\bibfnamefont {H.}~\bibnamefont {Pan}}, \ and\
  \bibinfo {author} {\bibfnamefont {H.-L.}\ \bibnamefont {Wang}},\ }\href@noop
  {} {\bibfield  {journal} {\bibinfo  {journal} {Physica B: Condensed Matter}\
  }\textbf {\bibinfo {volume} {511}},\ \bibinfo {pages} {80} (\bibinfo {year}
  {2017})}\BibitemShut {NoStop}%
\bibitem [{\citenamefont {Nguyen}\ \emph {et~al.}(2017)\citenamefont {Nguyen},
  \citenamefont {Lherbier},\ and\ \citenamefont
  {Charlier}}]{nguyen2017optical}%
  \BibitemOpen
  \bibfield  {author} {\bibinfo {author} {\bibfnamefont {V.~H.}\ \bibnamefont
  {Nguyen}}, \bibinfo {author} {\bibfnamefont {A.}~\bibnamefont {Lherbier}}, \
  and\ \bibinfo {author} {\bibfnamefont {J.-C.}\ \bibnamefont {Charlier}},\
  }\href@noop {} {\bibfield  {journal} {\bibinfo  {journal} {2D Materials}\
  }\textbf {\bibinfo {volume} {4}},\ \bibinfo {pages} {025041} (\bibinfo {year}
  {2017})}\BibitemShut {NoStop}%
\bibitem [{\citenamefont {Suzuura}\ and\ \citenamefont
  {Ando}(2002)}]{Suzuura02}%
  \BibitemOpen
  \bibfield  {author} {\bibinfo {author} {\bibfnamefont {H.}~\bibnamefont
  {Suzuura}}\ and\ \bibinfo {author} {\bibfnamefont {T.}~\bibnamefont {Ando}},\
  }\href {\doibase 10.1103/PhysRevB.65.235412} {\bibfield  {journal} {\bibinfo
  {journal} {Phys. Rev. B}\ }\textbf {\bibinfo {volume} {65}},\ \bibinfo
  {pages} {235412} (\bibinfo {year} {2002})}\BibitemShut {NoStop}%
\bibitem [{\citenamefont {Morpurgo}\ and\ \citenamefont
  {Guinea}(2006)}]{Morpurgo06}%
  \BibitemOpen
  \bibfield  {author} {\bibinfo {author} {\bibfnamefont {A.~F.}\ \bibnamefont
  {Morpurgo}}\ and\ \bibinfo {author} {\bibfnamefont {F.}~\bibnamefont
  {Guinea}},\ }\href {\doibase 10.1103/PhysRevLett.97.196804} {\bibfield
  {journal} {\bibinfo  {journal} {Phys. Rev. Lett.}\ }\textbf {\bibinfo
  {volume} {97}},\ \bibinfo {pages} {196804} (\bibinfo {year}
  {2006})}\BibitemShut {NoStop}%
\bibitem [{\citenamefont {Ma\~nes}(2007)}]{Manes07}%
  \BibitemOpen
  \bibfield  {author} {\bibinfo {author} {\bibfnamefont {J.~L.}\ \bibnamefont
  {Ma\~nes}},\ }\href {\doibase 10.1103/PhysRevB.76.045430} {\bibfield
  {journal} {\bibinfo  {journal} {Phys. Rev. B}\ }\textbf {\bibinfo {volume}
  {76}},\ \bibinfo {pages} {045430} (\bibinfo {year} {2007})}\BibitemShut
  {NoStop}%
\bibitem [{\citenamefont {Castro~Neto}\ \emph {et~al.}(2009)\citenamefont
  {Castro~Neto}, \citenamefont {Guinea}, \citenamefont {Peres}, \citenamefont
  {Novoselov},\ and\ \citenamefont {Geim}}]{Castro09}%
  \BibitemOpen
  \bibfield  {author} {\bibinfo {author} {\bibfnamefont {A.~H.}\ \bibnamefont
  {Castro~Neto}}, \bibinfo {author} {\bibfnamefont {F.}~\bibnamefont {Guinea}},
  \bibinfo {author} {\bibfnamefont {N.~M.~R.}\ \bibnamefont {Peres}}, \bibinfo
  {author} {\bibfnamefont {K.~S.}\ \bibnamefont {Novoselov}}, \ and\ \bibinfo
  {author} {\bibfnamefont {A.~K.}\ \bibnamefont {Geim}},\ }\href {\doibase
  10.1103/RevModPhys.81.109} {\bibfield  {journal} {\bibinfo  {journal} {Rev.
  Mod. Phys.}\ }\textbf {\bibinfo {volume} {81}},\ \bibinfo {pages} {109}
  (\bibinfo {year} {2009})}\BibitemShut {NoStop}%
\bibitem [{\citenamefont {Oliva-Leyva}\ and\ \citenamefont
  {Wang}(2017{\natexlab{a}})}]{oliva2017low-energy}%
  \BibitemOpen
  \bibfield  {author} {\bibinfo {author} {\bibfnamefont {M.}~\bibnamefont
  {Oliva-Leyva}}\ and\ \bibinfo {author} {\bibfnamefont {C.}~\bibnamefont
  {Wang}},\ }\href {http://stacks.iop.org/0953-8984/29/i=16/a=165301}
  {\bibfield  {journal} {\bibinfo  {journal} {Journal of Physics: Condensed
  Matter}\ }\textbf {\bibinfo {volume} {29}},\ \bibinfo {pages} {165301}
  (\bibinfo {year} {2017}{\natexlab{a}})}\BibitemShut {NoStop}%
\bibitem [{\citenamefont {Oliva-Leyva}\ and\ \citenamefont
  {Wang}(2017{\natexlab{b}})}]{oliva2017magento-optical}%
  \BibitemOpen
  \bibfield  {author} {\bibinfo {author} {\bibfnamefont {M.}~\bibnamefont
  {Oliva-Leyva}}\ and\ \bibinfo {author} {\bibfnamefont {C.}~\bibnamefont
  {Wang}},\ }\href {\doibase https://doi.org/10.1016/j.aop.2017.06.013}
  {\bibfield  {journal} {\bibinfo  {journal} {Annals of Physics}\ }\textbf
  {\bibinfo {volume} {384}},\ \bibinfo {pages} {61 } (\bibinfo {year}
  {2017}{\natexlab{b}})}\BibitemShut {NoStop}%
\bibitem [{\citenamefont {Vozmediano}\ \emph {et~al.}(2010)\citenamefont
  {Vozmediano}, \citenamefont {Katsnelson},\ and\ \citenamefont
  {Guinea}}]{Vozmediano10}%
  \BibitemOpen
  \bibfield  {author} {\bibinfo {author} {\bibfnamefont {M.}~\bibnamefont
  {Vozmediano}}, \bibinfo {author} {\bibfnamefont {M.}~\bibnamefont
  {Katsnelson}}, \ and\ \bibinfo {author} {\bibfnamefont {F.}~\bibnamefont
  {Guinea}},\ }\href {\doibase http://dx.doi.org/10.1016/j.physrep.2010.07.003}
  {\bibfield  {journal} {\bibinfo  {journal} {Physics Reports}\ }\textbf
  {\bibinfo {volume} {496}},\ \bibinfo {pages} {109 } (\bibinfo {year}
  {2010})}\BibitemShut {NoStop}%
\bibitem [{\citenamefont {Delplace}\ \emph {et~al.}(2011)\citenamefont
  {Delplace}, \citenamefont {Ullmo},\ and\ \citenamefont
  {Montambaux}}]{ZakPhase11}%
  \BibitemOpen
  \bibfield  {author} {\bibinfo {author} {\bibfnamefont {P.}~\bibnamefont
  {Delplace}}, \bibinfo {author} {\bibfnamefont {D.}~\bibnamefont {Ullmo}}, \
  and\ \bibinfo {author} {\bibfnamefont {G.}~\bibnamefont {Montambaux}},\
  }\href {\doibase 10.1103/PhysRevB.84.195452} {\bibfield  {journal} {\bibinfo
  {journal} {Phys. Rev. B}\ }\textbf {\bibinfo {volume} {84}},\ \bibinfo
  {pages} {195452} (\bibinfo {year} {2011})}\BibitemShut {NoStop}%
\bibitem [{\citenamefont {Roman-Taboada}\ and\ \citenamefont
  {Naumis}(2014)}]{Nosotros214}%
  \BibitemOpen
  \bibfield  {author} {\bibinfo {author} {\bibfnamefont {P.}~\bibnamefont
  {Roman-Taboada}}\ and\ \bibinfo {author} {\bibfnamefont {G.~G.}\ \bibnamefont
  {Naumis}},\ }\href {\doibase 10.1103/PhysRevB.90.195435} {\bibfield
  {journal} {\bibinfo  {journal} {Phys. Rev. B}\ }\textbf {\bibinfo {volume}
  {90}},\ \bibinfo {pages} {195435} (\bibinfo {year} {2014})}\BibitemShut
  {NoStop}%
\bibitem [{\citenamefont {Chou}\ and\ \citenamefont
  {Foster}(2014)}]{chou2014chalker}%
  \BibitemOpen
  \bibfield  {author} {\bibinfo {author} {\bibfnamefont {Y.-Z.}\ \bibnamefont
  {Chou}}\ and\ \bibinfo {author} {\bibfnamefont {M.~S.}\ \bibnamefont
  {Foster}},\ }\href@noop {} {\bibfield  {journal} {\bibinfo  {journal}
  {Physical Review B}\ }\textbf {\bibinfo {volume} {89}},\ \bibinfo {pages}
  {165136} (\bibinfo {year} {2014})}\BibitemShut {NoStop}%
\bibitem [{\citenamefont {Zyuzin}\ and\ \citenamefont
  {Zyuzin}(2015)}]{Zyuzin2015}%
  \BibitemOpen
  \bibfield  {author} {\bibinfo {author} {\bibfnamefont {A.~A.}\ \bibnamefont
  {Zyuzin}}\ and\ \bibinfo {author} {\bibfnamefont {V.~A.}\ \bibnamefont
  {Zyuzin}},\ }\href {\doibase 10.1134/S0021364015140143} {\bibfield  {journal}
  {\bibinfo  {journal} {JETP Letters}\ }\textbf {\bibinfo {volume} {102}},\
  \bibinfo {pages} {113} (\bibinfo {year} {2015})}\BibitemShut {NoStop}%
\bibitem [{\citenamefont {Guassi}\ \emph {et~al.}(2015)\citenamefont {Guassi},
  \citenamefont {Diniz}, \citenamefont {Sandler},\ and\ \citenamefont
  {Qu}}]{gaussi2015zero-field}%
  \BibitemOpen
  \bibfield  {author} {\bibinfo {author} {\bibfnamefont {M.~R.}\ \bibnamefont
  {Guassi}}, \bibinfo {author} {\bibfnamefont {G.~S.}\ \bibnamefont {Diniz}},
  \bibinfo {author} {\bibfnamefont {N.}~\bibnamefont {Sandler}}, \ and\
  \bibinfo {author} {\bibfnamefont {F.}~\bibnamefont {Qu}},\ }\href {\doibase
  10.1103/PhysRevB.92.075426} {\bibfield  {journal} {\bibinfo  {journal} {Phys.
  Rev. B}\ }\textbf {\bibinfo {volume} {92}},\ \bibinfo {pages} {075426}
  (\bibinfo {year} {2015})}\BibitemShut {NoStop}%
\bibitem [{\citenamefont {Mishra}\ \emph {et~al.}(2015)\citenamefont {Mishra},
  \citenamefont {Sarkar},\ and\ \citenamefont {Bandyopadhyay}}]{Mishra2015}%
  \BibitemOpen
  \bibfield  {author} {\bibinfo {author} {\bibfnamefont {T.}~\bibnamefont
  {Mishra}}, \bibinfo {author} {\bibfnamefont {T.~G.}\ \bibnamefont {Sarkar}},
  \ and\ \bibinfo {author} {\bibfnamefont {J.~N.}\ \bibnamefont
  {Bandyopadhyay}},\ }\href {\doibase 10.1140/epjb/e2015-60356-2} {\bibfield
  {journal} {\bibinfo  {journal} {The European Physical Journal B}\ }\textbf
  {\bibinfo {volume} {88}},\ \bibinfo {pages} {231} (\bibinfo {year}
  {2015})}\BibitemShut {NoStop}%
\bibitem [{\citenamefont {San-Jose}\ \emph {et~al.}(2015)\citenamefont
  {San-Jose}, \citenamefont {Lado}, \citenamefont {Aguado}, \citenamefont
  {Guinea},\ and\ \citenamefont {Fern\'andez-Rossier}}]{Majgraphene15}%
  \BibitemOpen
  \bibfield  {author} {\bibinfo {author} {\bibfnamefont {P.}~\bibnamefont
  {San-Jose}}, \bibinfo {author} {\bibfnamefont {J.~L.}\ \bibnamefont {Lado}},
  \bibinfo {author} {\bibfnamefont {R.}~\bibnamefont {Aguado}}, \bibinfo
  {author} {\bibfnamefont {F.}~\bibnamefont {Guinea}}, \ and\ \bibinfo {author}
  {\bibfnamefont {J.}~\bibnamefont {Fern\'andez-Rossier}},\ }\href {\doibase
  10.1103/PhysRevX.5.041042} {\bibfield  {journal} {\bibinfo  {journal} {Phys.
  Rev. X}\ }\textbf {\bibinfo {volume} {5}},\ \bibinfo {pages} {041042}
  (\bibinfo {year} {2015})}\BibitemShut {NoStop}%
\bibitem [{\citenamefont {Iorio}\ and\ \citenamefont
  {Pais}(2015)}]{iorio2015revisting}%
  \BibitemOpen
  \bibfield  {author} {\bibinfo {author} {\bibfnamefont {A.}~\bibnamefont
  {Iorio}}\ and\ \bibinfo {author} {\bibfnamefont {P.}~\bibnamefont {Pais}},\
  }\href {\doibase 10.1103/PhysRevD.92.125005} {\bibfield  {journal} {\bibinfo
  {journal} {Phys. Rev. D}\ }\textbf {\bibinfo {volume} {92}},\ \bibinfo
  {pages} {125005} (\bibinfo {year} {2015})}\BibitemShut {NoStop}%
\bibitem [{\citenamefont {Dal~Lago}\ and\ \citenamefont
  {Torres}(2015)}]{dal2015line}%
  \BibitemOpen
  \bibfield  {author} {\bibinfo {author} {\bibfnamefont {V.}~\bibnamefont
  {Dal~Lago}}\ and\ \bibinfo {author} {\bibfnamefont {L.~F.}\ \bibnamefont
  {Torres}},\ }\href@noop {} {\bibfield  {journal} {\bibinfo  {journal}
  {Journal of Physics: Condensed Matter}\ }\textbf {\bibinfo {volume} {27}},\
  \bibinfo {pages} {145303} (\bibinfo {year} {2015})}\BibitemShut {NoStop}%
\bibitem [{\citenamefont {Qu}\ \emph {et~al.}(2016)\citenamefont {Qu},
  \citenamefont {Diniz},\ and\ \citenamefont {Guassi}}]{qu2016electronic}%
  \BibitemOpen
  \bibfield  {author} {\bibinfo {author} {\bibfnamefont {F.}~\bibnamefont
  {Qu}}, \bibinfo {author} {\bibfnamefont {G.~S.}\ \bibnamefont {Diniz}}, \
  and\ \bibinfo {author} {\bibfnamefont {M.~R.}\ \bibnamefont {Guassi}},\ }in\
  \href {\doibase 10.5772/64493} {\emph {\bibinfo {booktitle} {Recent Advances
  in Graphene Research}}},\ \bibinfo {editor} {edited by\ \bibinfo {editor}
  {\bibfnamefont {P.~K.}\ \bibnamefont {Nayak}}}\ (\bibinfo  {publisher}
  {InTech},\ \bibinfo {address} {Rijeka},\ \bibinfo {year} {2016})\
  Chap.~\bibinfo {chapter} {03}\BibitemShut {NoStop}%
\bibitem [{\citenamefont {Frank}\ \emph {et~al.}(2017)\citenamefont {Frank},
  \citenamefont {H{\"o}gl}, \citenamefont {Gmitra}, \citenamefont {Kochan},\
  and\ \citenamefont {Fabian}}]{frank17}%
  \BibitemOpen
  \bibfield  {author} {\bibinfo {author} {\bibfnamefont {T.}~\bibnamefont
  {Frank}}, \bibinfo {author} {\bibfnamefont {P.}~\bibnamefont {H{\"o}gl}},
  \bibinfo {author} {\bibfnamefont {M.}~\bibnamefont {Gmitra}}, \bibinfo
  {author} {\bibfnamefont {D.}~\bibnamefont {Kochan}}, \ and\ \bibinfo {author}
  {\bibfnamefont {J.}~\bibnamefont {Fabian}},\ }\href@noop {} {\bibfield
  {journal} {\bibinfo  {journal} {arXiv preprint arXiv:1707.02124}\ } (\bibinfo
  {year} {2017})}\BibitemShut {NoStop}%
\bibitem [{\citenamefont {Cao}\ \emph {et~al.}(2017)\citenamefont {Cao},
  \citenamefont {Zhao},\ and\ \citenamefont {Louie}}]{cao2017topological}%
  \BibitemOpen
  \bibfield  {author} {\bibinfo {author} {\bibfnamefont {T.}~\bibnamefont
  {Cao}}, \bibinfo {author} {\bibfnamefont {F.}~\bibnamefont {Zhao}}, \ and\
  \bibinfo {author} {\bibfnamefont {S.~G.}\ \bibnamefont {Louie}},\ }\href
  {\doibase 10.1103/PhysRevLett.119.076401} {\bibfield  {journal} {\bibinfo
  {journal} {Phys. Rev. Lett.}\ }\textbf {\bibinfo {volume} {119}},\ \bibinfo
  {pages} {076401} (\bibinfo {year} {2017})}\BibitemShut {NoStop}%
\bibitem [{\citenamefont {Wu}\ \emph {et~al.}(2017)\citenamefont {Wu},
  \citenamefont {Shi}, \citenamefont {Sreejith},\ and\ \citenamefont
  {Liu}}]{wu2017fermionic}%
  \BibitemOpen
  \bibfield  {author} {\bibinfo {author} {\bibfnamefont {Y.-H.}\ \bibnamefont
  {Wu}}, \bibinfo {author} {\bibfnamefont {T.}~\bibnamefont {Shi}}, \bibinfo
  {author} {\bibfnamefont {G.}~\bibnamefont {Sreejith}}, \ and\ \bibinfo
  {author} {\bibfnamefont {Z.-X.}\ \bibnamefont {Liu}},\ }\href@noop {}
  {\bibfield  {journal} {\bibinfo  {journal} {Physical Review B}\ }\textbf
  {\bibinfo {volume} {96}},\ \bibinfo {pages} {085138} (\bibinfo {year}
  {2017})}\BibitemShut {NoStop}%
\bibitem [{\citenamefont {Wang}\ \emph
  {et~al.}(2017{\natexlab{a}})\citenamefont {Wang}, \citenamefont {Castro},\
  and\ \citenamefont {Lin}}]{wang2017strain}%
  \BibitemOpen
  \bibfield  {author} {\bibinfo {author} {\bibfnamefont {Z.-H.}\ \bibnamefont
  {Wang}}, \bibinfo {author} {\bibfnamefont {E.~V.}\ \bibnamefont {Castro}}, \
  and\ \bibinfo {author} {\bibfnamefont {H.-Q.}\ \bibnamefont {Lin}},\
  }\href@noop {} {\bibfield  {journal} {\bibinfo  {journal} {arXiv preprint
  arXiv:1708.00467}\ } (\bibinfo {year} {2017}{\natexlab{a}})}\BibitemShut
  {NoStop}%
\bibitem [{\citenamefont {Delplace}\ \emph
  {et~al.}(2013{\natexlab{a}})\citenamefont {Delplace}, \citenamefont
  {G\'omez-Le\'on},\ and\ \citenamefont {Platero}}]{Delplace13}%
  \BibitemOpen
  \bibfield  {author} {\bibinfo {author} {\bibfnamefont {P.}~\bibnamefont
  {Delplace}}, \bibinfo {author} {\bibfnamefont {A.}~\bibnamefont
  {G\'omez-Le\'on}}, \ and\ \bibinfo {author} {\bibfnamefont {G.}~\bibnamefont
  {Platero}},\ }\href {\doibase 10.1103/PhysRevB.88.245422} {\bibfield
  {journal} {\bibinfo  {journal} {Phys. Rev. B}\ }\textbf {\bibinfo {volume}
  {88}},\ \bibinfo {pages} {245422} (\bibinfo {year}
  {2013}{\natexlab{a}})}\BibitemShut {NoStop}%
\bibitem [{\citenamefont {Iadecola}\ \emph {et~al.}(2014)\citenamefont
  {Iadecola}, \citenamefont {Neupert},\ and\ \citenamefont
  {Chamon}}]{Thomas14}%
  \BibitemOpen
  \bibfield  {author} {\bibinfo {author} {\bibfnamefont {T.}~\bibnamefont
  {Iadecola}}, \bibinfo {author} {\bibfnamefont {T.}~\bibnamefont {Neupert}}, \
  and\ \bibinfo {author} {\bibfnamefont {C.}~\bibnamefont {Chamon}},\ }\href
  {\doibase 10.1103/PhysRevB.89.115425} {\bibfield  {journal} {\bibinfo
  {journal} {Phys. Rev. B}\ }\textbf {\bibinfo {volume} {89}},\ \bibinfo
  {pages} {115425} (\bibinfo {year} {2014})}\BibitemShut {NoStop}%
\bibitem [{\citenamefont {Gumbs}\ \emph {et~al.}(2014)\citenamefont {Gumbs},
  \citenamefont {Iurov}, \citenamefont {Huang},\ and\ \citenamefont
  {Zhemchuzhna}}]{Gumbs14}%
  \BibitemOpen
  \bibfield  {author} {\bibinfo {author} {\bibfnamefont {G.}~\bibnamefont
  {Gumbs}}, \bibinfo {author} {\bibfnamefont {A.}~\bibnamefont {Iurov}},
  \bibinfo {author} {\bibfnamefont {D.}~\bibnamefont {Huang}}, \ and\ \bibinfo
  {author} {\bibfnamefont {L.}~\bibnamefont {Zhemchuzhna}},\ }\href {\doibase
  10.1103/PhysRevB.89.241407} {\bibfield  {journal} {\bibinfo  {journal} {Phys.
  Rev. B}\ }\textbf {\bibinfo {volume} {89}},\ \bibinfo {pages} {241407}
  (\bibinfo {year} {2014})}\BibitemShut {NoStop}%
\bibitem [{\citenamefont {Perez-Piskunow}\ \emph {et~al.}(2014)\citenamefont
  {Perez-Piskunow}, \citenamefont {Usaj}, \citenamefont {Balseiro},\ and\
  \citenamefont {Torres}}]{perez2014floquet}%
  \BibitemOpen
  \bibfield  {author} {\bibinfo {author} {\bibfnamefont {P.}~\bibnamefont
  {Perez-Piskunow}}, \bibinfo {author} {\bibfnamefont {G.}~\bibnamefont
  {Usaj}}, \bibinfo {author} {\bibfnamefont {C.}~\bibnamefont {Balseiro}}, \
  and\ \bibinfo {author} {\bibfnamefont {L.~F.}\ \bibnamefont {Torres}},\
  }\href@noop {} {\bibfield  {journal} {\bibinfo  {journal} {Physical Review
  B}\ }\textbf {\bibinfo {volume} {89}},\ \bibinfo {pages} {121401} (\bibinfo
  {year} {2014})}\BibitemShut {NoStop}%
\bibitem [{\citenamefont {Usaj}\ \emph {et~al.}(2014)\citenamefont {Usaj},
  \citenamefont {Perez-Piskunow}, \citenamefont {Foa~Torres},\ and\
  \citenamefont {Balseiro}}]{Usaj14}%
  \BibitemOpen
  \bibfield  {author} {\bibinfo {author} {\bibfnamefont {G.}~\bibnamefont
  {Usaj}}, \bibinfo {author} {\bibfnamefont {P.~M.}\ \bibnamefont
  {Perez-Piskunow}}, \bibinfo {author} {\bibfnamefont {L.~E.~F.}\ \bibnamefont
  {Foa~Torres}}, \ and\ \bibinfo {author} {\bibfnamefont {C.~A.}\ \bibnamefont
  {Balseiro}},\ }\href {\doibase 10.1103/PhysRevB.90.115423} {\bibfield
  {journal} {\bibinfo  {journal} {Phys. Rev. B}\ }\textbf {\bibinfo {volume}
  {90}},\ \bibinfo {pages} {115423} (\bibinfo {year} {2014})}\BibitemShut
  {NoStop}%
\bibitem [{\citenamefont {Gavensky}\ \emph {et~al.}(2016)\citenamefont
  {Gavensky}, \citenamefont {Usaj},\ and\ \citenamefont
  {Balseiro}}]{gavensky16}%
  \BibitemOpen
  \bibfield  {author} {\bibinfo {author} {\bibfnamefont {L.~P.}\ \bibnamefont
  {Gavensky}}, \bibinfo {author} {\bibfnamefont {G.}~\bibnamefont {Usaj}}, \
  and\ \bibinfo {author} {\bibfnamefont {C.}~\bibnamefont {Balseiro}},\
  }\href@noop {} {\bibfield  {journal} {\bibinfo  {journal} {Scientific
  reports}\ }\textbf {\bibinfo {volume} {6}},\ \bibinfo {pages} {36577}
  (\bibinfo {year} {2016})}\BibitemShut {NoStop}%
\bibitem [{\citenamefont {Manghi}\ \emph {et~al.}(2017)\citenamefont {Manghi},
  \citenamefont {Puviani},\ and\ \citenamefont {Lenzini}}]{manghi17}%
  \BibitemOpen
  \bibfield  {author} {\bibinfo {author} {\bibfnamefont {F.}~\bibnamefont
  {Manghi}}, \bibinfo {author} {\bibfnamefont {M.}~\bibnamefont {Puviani}}, \
  and\ \bibinfo {author} {\bibfnamefont {F.}~\bibnamefont {Lenzini}},\
  }\href@noop {} {\bibfield  {journal} {\bibinfo  {journal} {arXiv preprint
  arXiv:1702.06349}\ } (\bibinfo {year} {2017})}\BibitemShut {NoStop}%
\bibitem [{\citenamefont {Lago}\ \emph {et~al.}(2017)\citenamefont {Lago},
  \citenamefont {Morell},\ and\ \citenamefont {Torres}}]{lago17}%
  \BibitemOpen
  \bibfield  {author} {\bibinfo {author} {\bibfnamefont {V.~D.}\ \bibnamefont
  {Lago}}, \bibinfo {author} {\bibfnamefont {E.~S.}\ \bibnamefont {Morell}}, \
  and\ \bibinfo {author} {\bibfnamefont {L.~E.}\ \bibnamefont {Torres}},\
  }\href@noop {} {\bibfield  {journal} {\bibinfo  {journal} {arXiv preprint
  arXiv:1708.03304}\ } (\bibinfo {year} {2017})}\BibitemShut {NoStop}%
\bibitem [{\citenamefont {Roman-Taboada}\ and\ \citenamefont
  {Naumis}(2017)}]{roman2017topological}%
  \BibitemOpen
  \bibfield  {author} {\bibinfo {author} {\bibfnamefont {P.}~\bibnamefont
  {Roman-Taboada}}\ and\ \bibinfo {author} {\bibfnamefont {G.~G.}\ \bibnamefont
  {Naumis}},\ }\href@noop {} {\bibfield  {journal} {\bibinfo  {journal}
  {Physical Review B}\ }\textbf {\bibinfo {volume} {95}},\ \bibinfo {pages}
  {115440} (\bibinfo {year} {2017})}\BibitemShut {NoStop}%
\bibitem [{\citenamefont {Novoselov}\ \emph {et~al.}(2005)\citenamefont
  {Novoselov}, \citenamefont {Geim}, \citenamefont {Morozov}, \citenamefont
  {Jiang}, \citenamefont {Katsnelson}, \citenamefont {Grigorieva},
  \citenamefont {Dubonos},\ and\ \citenamefont {Firsov}}]{Novoselov05}%
  \BibitemOpen
  \bibfield  {author} {\bibinfo {author} {\bibfnamefont {K.~S.}\ \bibnamefont
  {Novoselov}}, \bibinfo {author} {\bibfnamefont {A.~K.}\ \bibnamefont {Geim}},
  \bibinfo {author} {\bibfnamefont {S.~V.}\ \bibnamefont {Morozov}}, \bibinfo
  {author} {\bibfnamefont {D.}~\bibnamefont {Jiang}}, \bibinfo {author}
  {\bibfnamefont {M.~I.}\ \bibnamefont {Katsnelson}}, \bibinfo {author}
  {\bibnamefont {Grigorieva}}, \bibinfo {author} {\bibfnamefont {S.~V.}\
  \bibnamefont {Dubonos}}, \ and\ \bibinfo {author} {\bibfnamefont {A.~A.}\
  \bibnamefont {Firsov}},\ }\href {\doibase 10.1038/nature04233} {\bibfield
  {journal} {\bibinfo  {journal} {Nature}\ }\textbf {\bibinfo {volume} {438}},\
  \bibinfo {pages} {197} (\bibinfo {year} {2005})}\BibitemShut {NoStop}%
\bibitem [{\citenamefont {Delplace}\ \emph
  {et~al.}(2013{\natexlab{b}})\citenamefont {Delplace}, \citenamefont
  {G\'omez-Le\'on},\ and\ \citenamefont {Platero}}]{delplace2013merging}%
  \BibitemOpen
  \bibfield  {author} {\bibinfo {author} {\bibfnamefont {P.}~\bibnamefont
  {Delplace}}, \bibinfo {author} {\bibfnamefont {A.}~\bibnamefont
  {G\'omez-Le\'on}}, \ and\ \bibinfo {author} {\bibfnamefont {G.}~\bibnamefont
  {Platero}},\ }\href {\doibase 10.1103/PhysRevB.88.245422} {\bibfield
  {journal} {\bibinfo  {journal} {Phys. Rev. B}\ }\textbf {\bibinfo {volume}
  {88}},\ \bibinfo {pages} {245422} (\bibinfo {year}
  {2013}{\natexlab{b}})}\BibitemShut {NoStop}%
\bibitem [{\citenamefont {L\'opez-Rodr\'{\i}guez}\ and\ \citenamefont
  {Naumis}(2008)}]{Lopez2008}%
  \BibitemOpen
  \bibfield  {author} {\bibinfo {author} {\bibfnamefont {F.~J.}\ \bibnamefont
  {L\'opez-Rodr\'{\i}guez}}\ and\ \bibinfo {author} {\bibfnamefont {G.~G.}\
  \bibnamefont {Naumis}},\ }\href {\doibase 10.1103/PhysRevB.78.201406}
  {\bibfield  {journal} {\bibinfo  {journal} {Phys. Rev. B}\ }\textbf {\bibinfo
  {volume} {78}},\ \bibinfo {pages} {201406} (\bibinfo {year}
  {2008})}\BibitemShut {NoStop}%
\bibitem [{\citenamefont {L\'opez-Rodríguez}\ and\ \citenamefont
  {Naumis}(2010)}]{Lopez2010}%
  \BibitemOpen
  \bibfield  {author} {\bibinfo {author} {\bibfnamefont {F.}~\bibnamefont
  {L\'opez-Rodríguez}}\ and\ \bibinfo {author} {\bibfnamefont
  {G.}~\bibnamefont {Naumis}},\ }\href {\doibase 10.1080/14786431003757794}
  {\bibfield  {journal} {\bibinfo  {journal} {Philosophical Magazine}\ }\textbf
  {\bibinfo {volume} {90}},\ \bibinfo {pages} {2977} (\bibinfo {year}
  {2010})},\ \Eprint
  {http://arxiv.org/abs/http://dx.doi.org/10.1080/14786431003757794}
  {http://dx.doi.org/10.1080/14786431003757794} \BibitemShut {NoStop}%
\bibitem [{\citenamefont {Heikkil{\"a}}\ \emph {et~al.}(2011)\citenamefont
  {Heikkil{\"a}}, \citenamefont {Kopnin},\ and\ \citenamefont
  {Volovik}}]{Volovik2011}%
  \BibitemOpen
  \bibfield  {author} {\bibinfo {author} {\bibfnamefont {T.~T.}\ \bibnamefont
  {Heikkil{\"a}}}, \bibinfo {author} {\bibfnamefont {N.~B.}\ \bibnamefont
  {Kopnin}}, \ and\ \bibinfo {author} {\bibfnamefont {G.~E.}\ \bibnamefont
  {Volovik}},\ }\href {\doibase 10.1134/S0021364011150045} {\bibfield
  {journal} {\bibinfo  {journal} {JETP Letters}\ }\textbf {\bibinfo {volume}
  {94}},\ \bibinfo {pages} {233} (\bibinfo {year} {2011})}\BibitemShut
  {NoStop}%
\bibitem [{\citenamefont {Volovik}(2013)}]{Volovik2013}%
  \BibitemOpen
  \bibfield  {author} {\bibinfo {author} {\bibfnamefont {G.~E.}\ \bibnamefont
  {Volovik}},\ }\href {\doibase 10.1007/s10948-013-2221-5} {\bibfield
  {journal} {\bibinfo  {journal} {Journal of Superconductivity and Novel
  Magnetism}\ }\textbf {\bibinfo {volume} {26}},\ \bibinfo {pages} {2887}
  (\bibinfo {year} {2013})}\BibitemShut {NoStop}%
\bibitem [{\citenamefont {Bomantara}\ \emph {et~al.}(2016)\citenamefont
  {Bomantara}, \citenamefont {Raghava}, \citenamefont {Zhou},\ and\
  \citenamefont {Gong}}]{KHM16}%
  \BibitemOpen
  \bibfield  {author} {\bibinfo {author} {\bibfnamefont {R.~W.}\ \bibnamefont
  {Bomantara}}, \bibinfo {author} {\bibfnamefont {G.~N.}\ \bibnamefont
  {Raghava}}, \bibinfo {author} {\bibfnamefont {L.}~\bibnamefont {Zhou}}, \
  and\ \bibinfo {author} {\bibfnamefont {J.}~\bibnamefont {Gong}},\ }\href
  {\doibase 10.1103/PhysRevE.93.022209} {\bibfield  {journal} {\bibinfo
  {journal} {Phys. Rev. E}\ }\textbf {\bibinfo {volume} {93}},\ \bibinfo
  {pages} {022209} (\bibinfo {year} {2016})}\BibitemShut {NoStop}%
\bibitem [{\citenamefont {Wang}\ \emph
  {et~al.}(2017{\natexlab{b}})\citenamefont {Wang}, \citenamefont {Chen},
  \citenamefont {Bomantara}, \citenamefont {Gong},\ and\ \citenamefont
  {Xing}}]{wang2017linenodes}%
  \BibitemOpen
  \bibfield  {author} {\bibinfo {author} {\bibfnamefont {H.-Q.}\ \bibnamefont
  {Wang}}, \bibinfo {author} {\bibfnamefont {M.~N.}\ \bibnamefont {Chen}},
  \bibinfo {author} {\bibfnamefont {R.~W.}\ \bibnamefont {Bomantara}}, \bibinfo
  {author} {\bibfnamefont {J.}~\bibnamefont {Gong}}, \ and\ \bibinfo {author}
  {\bibfnamefont {D.~Y.}\ \bibnamefont {Xing}},\ }\href {\doibase
  10.1103/PhysRevB.95.075136} {\bibfield  {journal} {\bibinfo  {journal} {Phys.
  Rev. B}\ }\textbf {\bibinfo {volume} {95}},\ \bibinfo {pages} {075136}
  (\bibinfo {year} {2017}{\natexlab{b}})}\BibitemShut {NoStop}%
\bibitem [{\citenamefont {Bai}\ \emph {et~al.}(2014)\citenamefont {Bai},
  \citenamefont {Zhou}, \citenamefont {Zheng}, \citenamefont {Meng},
  \citenamefont {Peng}, \citenamefont {Liu}, \citenamefont {Nie},\ and\
  \citenamefont {He}}]{bai2014creating}%
  \BibitemOpen
  \bibfield  {author} {\bibinfo {author} {\bibfnamefont {K.-K.}\ \bibnamefont
  {Bai}}, \bibinfo {author} {\bibfnamefont {Y.}~\bibnamefont {Zhou}}, \bibinfo
  {author} {\bibfnamefont {H.}~\bibnamefont {Zheng}}, \bibinfo {author}
  {\bibfnamefont {L.}~\bibnamefont {Meng}}, \bibinfo {author} {\bibfnamefont
  {H.}~\bibnamefont {Peng}}, \bibinfo {author} {\bibfnamefont {Z.}~\bibnamefont
  {Liu}}, \bibinfo {author} {\bibfnamefont {J.-C.}\ \bibnamefont {Nie}}, \ and\
  \bibinfo {author} {\bibfnamefont {L.}~\bibnamefont {He}},\ }\href {\doibase
  10.1103/PhysRevLett.113.086102} {\bibfield  {journal} {\bibinfo  {journal}
  {Phys. Rev. Lett.}\ }\textbf {\bibinfo {volume} {113}},\ \bibinfo {pages}
  {086102} (\bibinfo {year} {2014})}\BibitemShut {NoStop}%
\bibitem [{\citenamefont {Agarwala}\ \emph {et~al.}(2016)\citenamefont
  {Agarwala}, \citenamefont {Bhattacharya}, \citenamefont {Dutta},\ and\
  \citenamefont {Sen}}]{agarwala2016effects}%
  \BibitemOpen
  \bibfield  {author} {\bibinfo {author} {\bibfnamefont {A.}~\bibnamefont
  {Agarwala}}, \bibinfo {author} {\bibfnamefont {U.}~\bibnamefont
  {Bhattacharya}}, \bibinfo {author} {\bibfnamefont {A.}~\bibnamefont {Dutta}},
  \ and\ \bibinfo {author} {\bibfnamefont {D.}~\bibnamefont {Sen}},\ }\href
  {\doibase 10.1103/PhysRevB.93.174301} {\bibfield  {journal} {\bibinfo
  {journal} {Phys. Rev. B}\ }\textbf {\bibinfo {volume} {93}},\ \bibinfo
  {pages} {174301} (\bibinfo {year} {2016})}\BibitemShut {NoStop}%
\bibitem [{\citenamefont {Uehlinger}\ \emph {et~al.}(2013)\citenamefont
  {Uehlinger}, \citenamefont {Jotzu}, \citenamefont {Messer}, \citenamefont
  {Greif}, \citenamefont {Hofstetter}, \citenamefont {Bissbort},\ and\
  \citenamefont {Esslinger}}]{Uehlinger13}%
  \BibitemOpen
  \bibfield  {author} {\bibinfo {author} {\bibfnamefont {T.}~\bibnamefont
  {Uehlinger}}, \bibinfo {author} {\bibfnamefont {G.}~\bibnamefont {Jotzu}},
  \bibinfo {author} {\bibfnamefont {M.}~\bibnamefont {Messer}}, \bibinfo
  {author} {\bibfnamefont {D.}~\bibnamefont {Greif}}, \bibinfo {author}
  {\bibfnamefont {W.}~\bibnamefont {Hofstetter}}, \bibinfo {author}
  {\bibfnamefont {U.}~\bibnamefont {Bissbort}}, \ and\ \bibinfo {author}
  {\bibfnamefont {T.}~\bibnamefont {Esslinger}},\ }\href {\doibase
  10.1103/PhysRevLett.111.185307} {\bibfield  {journal} {\bibinfo  {journal}
  {Phys. Rev. Lett.}\ }\textbf {\bibinfo {volume} {111}},\ \bibinfo {pages}
  {185307} (\bibinfo {year} {2013})}\BibitemShut {NoStop}%
\bibitem [{\citenamefont {Feng}\ \emph {et~al.}(2013)\citenamefont {Feng},
  \citenamefont {Dan-Wei},\ and\ \citenamefont {Shi-Liang}}]{mei2013graphene}%
  \BibitemOpen
  \bibfield  {author} {\bibinfo {author} {\bibfnamefont {M.}~\bibnamefont
  {Feng}}, \bibinfo {author} {\bibfnamefont {Z.}~\bibnamefont {Dan-Wei}}, \
  and\ \bibinfo {author} {\bibfnamefont {Z.}~\bibnamefont {Shi-Liang}},\ }\href
  {http://stacks.iop.org/1674-1056/22/i=11/a=116106} {\bibfield  {journal}
  {\bibinfo  {journal} {Chinese Physics B}\ }\textbf {\bibinfo {volume} {22}},\
  \bibinfo {pages} {116106} (\bibinfo {year} {2013})}\BibitemShut {NoStop}%
\bibitem [{\citenamefont {Feilhauer}\ \emph {et~al.}(2015)\citenamefont
  {Feilhauer}, \citenamefont {Apel},\ and\ \citenamefont
  {Schweitzer}}]{Feilhauer15}%
  \BibitemOpen
  \bibfield  {author} {\bibinfo {author} {\bibfnamefont {J.}~\bibnamefont
  {Feilhauer}}, \bibinfo {author} {\bibfnamefont {W.}~\bibnamefont {Apel}}, \
  and\ \bibinfo {author} {\bibfnamefont {L.}~\bibnamefont {Schweitzer}},\
  }\href {\doibase 10.1103/PhysRevB.92.245424} {\bibfield  {journal} {\bibinfo
  {journal} {Phys. Rev. B}\ }\textbf {\bibinfo {volume} {92}},\ \bibinfo
  {pages} {245424} (\bibinfo {year} {2015})}\BibitemShut {NoStop}%
\bibitem [{\citenamefont {Weinberg}\ \emph {et~al.}(2016)\citenamefont
  {Weinberg}, \citenamefont {Staarmann}, \citenamefont {Ölschläger},
  \citenamefont {Simonet},\ and\ \citenamefont {Sengstock}}]{weinberg16}%
  \BibitemOpen
  \bibfield  {author} {\bibinfo {author} {\bibfnamefont {M.}~\bibnamefont
  {Weinberg}}, \bibinfo {author} {\bibfnamefont {C.}~\bibnamefont {Staarmann}},
  \bibinfo {author} {\bibfnamefont {C.}~\bibnamefont {Ölschläger}}, \bibinfo
  {author} {\bibfnamefont {J.}~\bibnamefont {Simonet}}, \ and\ \bibinfo
  {author} {\bibfnamefont {K.}~\bibnamefont {Sengstock}},\ }\href
  {http://stacks.iop.org/2053-1583/3/i=2/a=024005} {\bibfield  {journal}
  {\bibinfo  {journal} {2D Materials}\ }\textbf {\bibinfo {volume} {3}},\
  \bibinfo {pages} {024005} (\bibinfo {year} {2016})}\BibitemShut {NoStop}%
\bibitem [{\citenamefont {Fl{\"a}schner}\ \emph {et~al.}(2016)\citenamefont
  {Fl{\"a}schner}, \citenamefont {Rem}, \citenamefont {Tarnowski},
  \citenamefont {Vogel}, \citenamefont {L{\"u}hmann}, \citenamefont
  {Sengstock},\ and\ \citenamefont {Weitenberg}}]{flaschner2016experimental}%
  \BibitemOpen
  \bibfield  {author} {\bibinfo {author} {\bibfnamefont {N.}~\bibnamefont
  {Fl{\"a}schner}}, \bibinfo {author} {\bibfnamefont {B.~S.}\ \bibnamefont
  {Rem}}, \bibinfo {author} {\bibfnamefont {M.}~\bibnamefont {Tarnowski}},
  \bibinfo {author} {\bibfnamefont {D.}~\bibnamefont {Vogel}}, \bibinfo
  {author} {\bibfnamefont {D.-S.}\ \bibnamefont {L{\"u}hmann}}, \bibinfo
  {author} {\bibfnamefont {K.}~\bibnamefont {Sengstock}}, \ and\ \bibinfo
  {author} {\bibfnamefont {C.}~\bibnamefont {Weitenberg}},\ }\href {\doibase
  10.1126/science.aad4568} {\bibfield  {journal} {\bibinfo  {journal}
  {Science}\ }\textbf {\bibinfo {volume} {352}},\ \bibinfo {pages} {1091}
  (\bibinfo {year} {2016})},\ \Eprint
  {http://arxiv.org/abs/http://science.sciencemag.org/content/352/6289/1091.full.pdf}
  {http://science.sciencemag.org/content/352/6289/1091.full.pdf} \BibitemShut
  {NoStop}%
\bibitem [{\citenamefont {Dautova}\ \emph {et~al.}(2017)\citenamefont
  {Dautova}, \citenamefont {Shytov}, \citenamefont {Hooper}, \citenamefont
  {Sambles},\ and\ \citenamefont {Hibbins}}]{dautova2017gapless}%
  \BibitemOpen
  \bibfield  {author} {\bibinfo {author} {\bibfnamefont {Y.~N.}\ \bibnamefont
  {Dautova}}, \bibinfo {author} {\bibfnamefont {A.~V.}\ \bibnamefont {Shytov}},
  \bibinfo {author} {\bibfnamefont {I.~R.}\ \bibnamefont {Hooper}}, \bibinfo
  {author} {\bibfnamefont {J.~R.}\ \bibnamefont {Sambles}}, \ and\ \bibinfo
  {author} {\bibfnamefont {A.~P.}\ \bibnamefont {Hibbins}},\ }\href@noop {}
  {\bibfield  {journal} {\bibinfo  {journal} {Applied Physics Letters}\
  }\textbf {\bibinfo {volume} {110}},\ \bibinfo {pages} {261605} (\bibinfo
  {year} {2017})}\BibitemShut {NoStop}%
\bibitem [{\citenamefont {Rudner}\ \emph {et~al.}(2013)\citenamefont {Rudner},
  \citenamefont {Lindner}, \citenamefont {Berg},\ and\ \citenamefont
  {Levin}}]{rudner2013anomalous}%
  \BibitemOpen
  \bibfield  {author} {\bibinfo {author} {\bibfnamefont {M.~S.}\ \bibnamefont
  {Rudner}}, \bibinfo {author} {\bibfnamefont {N.~H.}\ \bibnamefont {Lindner}},
  \bibinfo {author} {\bibfnamefont {E.}~\bibnamefont {Berg}}, \ and\ \bibinfo
  {author} {\bibfnamefont {M.}~\bibnamefont {Levin}},\ }\href {\doibase
  10.1103/PhysRevX.3.031005} {\bibfield  {journal} {\bibinfo  {journal} {Phys.
  Rev. X}\ }\textbf {\bibinfo {volume} {3}},\ \bibinfo {pages} {031005}
  (\bibinfo {year} {2013})}\BibitemShut {NoStop}%
\bibitem [{\citenamefont {Winkler}\ and\ \citenamefont
  {Deshpande}(2017)}]{Winkler17}%
  \BibitemOpen
  \bibfield  {author} {\bibinfo {author} {\bibfnamefont {R.}~\bibnamefont
  {Winkler}}\ and\ \bibinfo {author} {\bibfnamefont {H.}~\bibnamefont
  {Deshpande}},\ }\href {\doibase 10.1103/PhysRevB.95.235312} {\bibfield
  {journal} {\bibinfo  {journal} {Phys. Rev. B}\ }\textbf {\bibinfo {volume}
  {95}},\ \bibinfo {pages} {235312} (\bibinfo {year} {2017})}\BibitemShut
  {NoStop}%
\bibitem [{\citenamefont {L\'opez}\ \emph {et~al.}(1993)\citenamefont
  {L\'opez}, \citenamefont {Naumis},\ and\ \citenamefont
  {Arag\'on}}]{Naumis1993}%
  \BibitemOpen
  \bibfield  {author} {\bibinfo {author} {\bibfnamefont {J.~C.}\ \bibnamefont
  {L\'opez}}, \bibinfo {author} {\bibfnamefont {G.}~\bibnamefont {Naumis}}, \
  and\ \bibinfo {author} {\bibfnamefont {J.~L.}\ \bibnamefont {Arag\'on}},\
  }\href {\doibase 10.1103/PhysRevB.48.12459} {\bibfield  {journal} {\bibinfo
  {journal} {Phys. Rev. B}\ }\textbf {\bibinfo {volume} {48}},\ \bibinfo
  {pages} {12459} (\bibinfo {year} {1993})}\BibitemShut {NoStop}%
\bibitem [{\citenamefont {Naumis}(1999)}]{Naumis1999}%
  \BibitemOpen
  \bibfield  {author} {\bibinfo {author} {\bibfnamefont {G.~G.}\ \bibnamefont
  {Naumis}},\ }\href {\doibase 10.1103/PhysRevB.59.11315} {\bibfield  {journal}
  {\bibinfo  {journal} {Phys. Rev. B}\ }\textbf {\bibinfo {volume} {59}},\
  \bibinfo {pages} {11315} (\bibinfo {year} {1999})}\BibitemShut {NoStop}%
\bibitem [{\citenamefont {Naumis}\ \emph {et~al.}(1999)\citenamefont {Naumis},
  \citenamefont {Wang}, \citenamefont {Thorpe},\ and\ \citenamefont
  {Barrio}}]{NaumisThorpe1999}%
  \BibitemOpen
  \bibfield  {author} {\bibinfo {author} {\bibfnamefont {G.~G.}\ \bibnamefont
  {Naumis}}, \bibinfo {author} {\bibfnamefont {C.}~\bibnamefont {Wang}},
  \bibinfo {author} {\bibfnamefont {M.~F.}\ \bibnamefont {Thorpe}}, \ and\
  \bibinfo {author} {\bibfnamefont {R.~A.}\ \bibnamefont {Barrio}},\ }\href
  {\doibase 10.1103/PhysRevB.59.14302} {\bibfield  {journal} {\bibinfo
  {journal} {Phys. Rev. B}\ }\textbf {\bibinfo {volume} {59}},\ \bibinfo
  {pages} {14302} (\bibinfo {year} {1999})}\BibitemShut {NoStop}%
\bibitem [{\citenamefont {Satija}\ and\ \citenamefont
  {Naumis}(2013)}]{Satija2013}%
  \BibitemOpen
  \bibfield  {author} {\bibinfo {author} {\bibfnamefont {I.~I.}\ \bibnamefont
  {Satija}}\ and\ \bibinfo {author} {\bibfnamefont {G.~G.}\ \bibnamefont
  {Naumis}},\ }\href {\doibase 10.1103/PhysRevB.88.054204} {\bibfield
  {journal} {\bibinfo  {journal} {Phys. Rev. B}\ }\textbf {\bibinfo {volume}
  {88}},\ \bibinfo {pages} {054204} (\bibinfo {year} {2013})}\BibitemShut
  {NoStop}%
\bibitem [{\citenamefont {Pereira}\ \emph {et~al.}(2009)\citenamefont
  {Pereira}, \citenamefont {Castro~Neto},\ and\ \citenamefont
  {Peres}}]{Pereira09}%
  \BibitemOpen
  \bibfield  {author} {\bibinfo {author} {\bibfnamefont {V.~M.}\ \bibnamefont
  {Pereira}}, \bibinfo {author} {\bibfnamefont {A.~H.}\ \bibnamefont
  {Castro~Neto}}, \ and\ \bibinfo {author} {\bibfnamefont {N.~M.~R.}\
  \bibnamefont {Peres}},\ }\href {\doibase 10.1103/PhysRevB.80.045401}
  {\bibfield  {journal} {\bibinfo  {journal} {Phys. Rev. B}\ }\textbf {\bibinfo
  {volume} {80}},\ \bibinfo {pages} {045401} (\bibinfo {year}
  {2009})}\BibitemShut {NoStop}%
\bibitem [{\citenamefont {Yoshimura}\ \emph {et~al.}(2014)\citenamefont
  {Yoshimura}, \citenamefont {Imura}, \citenamefont {Fukui},\ and\
  \citenamefont {Hatsugai}}]{Yoshimura14}%
  \BibitemOpen
  \bibfield  {author} {\bibinfo {author} {\bibfnamefont {Y.}~\bibnamefont
  {Yoshimura}}, \bibinfo {author} {\bibfnamefont {K.-I.}\ \bibnamefont
  {Imura}}, \bibinfo {author} {\bibfnamefont {T.}~\bibnamefont {Fukui}}, \ and\
  \bibinfo {author} {\bibfnamefont {Y.}~\bibnamefont {Hatsugai}},\ }\href
  {\doibase 10.1103/PhysRevB.90.155443} {\bibfield  {journal} {\bibinfo
  {journal} {Phys. Rev. B}\ }\textbf {\bibinfo {volume} {90}},\ \bibinfo
  {pages} {155443} (\bibinfo {year} {2014})}\BibitemShut {NoStop}%
\bibitem [{\citenamefont {Ho}\ and\ \citenamefont {Gong}(2014)}]{Ho14}%
  \BibitemOpen
  \bibfield  {author} {\bibinfo {author} {\bibfnamefont {D.~Y.~H.}\
  \bibnamefont {Ho}}\ and\ \bibinfo {author} {\bibfnamefont {J.}~\bibnamefont
  {Gong}},\ }\href {\doibase 10.1103/PhysRevB.90.195419} {\bibfield  {journal}
  {\bibinfo  {journal} {Phys. Rev. B}\ }\textbf {\bibinfo {volume} {90}},\
  \bibinfo {pages} {195419} (\bibinfo {year} {2014})}\BibitemShut {NoStop}%
\bibitem [{\citenamefont {Sedlmayr}\ \emph {et~al.}(2015)\citenamefont
  {Sedlmayr}, \citenamefont {Aguiar-Hualde},\ and\ \citenamefont
  {Bena}}]{WeakTop15}%
  \BibitemOpen
  \bibfield  {author} {\bibinfo {author} {\bibfnamefont {N.}~\bibnamefont
  {Sedlmayr}}, \bibinfo {author} {\bibfnamefont {J.~M.}\ \bibnamefont
  {Aguiar-Hualde}}, \ and\ \bibinfo {author} {\bibfnamefont {C.}~\bibnamefont
  {Bena}},\ }\href {\doibase 10.1103/PhysRevB.91.115415} {\bibfield  {journal}
  {\bibinfo  {journal} {Phys. Rev. B}\ }\textbf {\bibinfo {volume} {91}},\
  \bibinfo {pages} {115415} (\bibinfo {year} {2015})}\BibitemShut {NoStop}%
\bibitem [{\citenamefont {Koghee}\ \emph {et~al.}(2012)\citenamefont {Koghee},
  \citenamefont {Lim}, \citenamefont {Goerbig},\ and\ \citenamefont
  {Smith}}]{koghee2012merging}%
  \BibitemOpen
  \bibfield  {author} {\bibinfo {author} {\bibfnamefont {S.}~\bibnamefont
  {Koghee}}, \bibinfo {author} {\bibfnamefont {L.-K.}\ \bibnamefont {Lim}},
  \bibinfo {author} {\bibfnamefont {M.~O.}\ \bibnamefont {Goerbig}}, \ and\
  \bibinfo {author} {\bibfnamefont {C.~M.}\ \bibnamefont {Smith}},\ }\href
  {\doibase 10.1103/PhysRevA.85.023637} {\bibfield  {journal} {\bibinfo
  {journal} {Phys. Rev. A}\ }\textbf {\bibinfo {volume} {85}},\ \bibinfo
  {pages} {023637} (\bibinfo {year} {2012})}\BibitemShut {NoStop}%
\bibitem [{\citenamefont {Zheng}\ and\ \citenamefont
  {Zhai}(2014)}]{zheng2014floquet}%
  \BibitemOpen
  \bibfield  {author} {\bibinfo {author} {\bibfnamefont {W.}~\bibnamefont
  {Zheng}}\ and\ \bibinfo {author} {\bibfnamefont {H.}~\bibnamefont {Zhai}},\
  }\href {\doibase 10.1103/PhysRevA.89.061603} {\bibfield  {journal} {\bibinfo
  {journal} {Phys. Rev. A}\ }\textbf {\bibinfo {volume} {89}},\ \bibinfo
  {pages} {061603} (\bibinfo {year} {2014})}\BibitemShut {NoStop}%
\end{thebibliography}%
\end{document}